\documentclass[prb,showpacs,groupedaddress,twocolumn]{revtex4}
\usepackage{amsmath}
\usepackage{mathrsfs}
\usepackage{txfonts}
\usepackage{amssymb}
\usepackage{graphicx}
\usepackage{hyperref}
\usepackage{ulem}
 \usepackage{overpic}
 \usepackage{psfrag}
\begin{document}

\title{ Quantum fidelity for degenerate groundstates in quantum phase transitions}
\author{Yao Heng Su}
\affiliation{Centre for Modern Physics and Department of Physics,
Chongqing University, Chongqing 400044, China}

\author{Bing-Quan Hu}
\affiliation{Centre for Modern Physics and Department of Physics,
Chongqing University, Chongqing 400044, China}

\author{Sheng-Hao Li}
\affiliation{Centre for Modern Physics and Department of Physics,
Chongqing University, Chongqing 400044, China}

\author{Sam Young Cho}
\email{sycho@cqu.edu.cn}
 \affiliation{Centre for Modern Physics and
Department of Physics, Chongqing University, Chongqing 400044,
China}

%
%\date{\today}
%
%
%
\begin{abstract}
 Spontaneous symmetry breaking mechanism in quantum phase
 transitions manifests the existence of degenerate groundstates in broken symmetry
 phases.
 To detect such degenerate groundstates,
 we introduce a quantum fidelity as an overlap measurement
 between system groundstates and an arbitrary
 reference state.
 This quantum fidelity is shown a multiple bifurcation as an indicator
 of quantum phase transitions, without knowing any detailed broken symmetry,
 between a broken symmetry phase and symmetry
 phases
 as well as between a broken symmetry phase and other broken symmetry
 phases, when a system parameter crosses its critical value (i.e., multiple
 bifurcation point). Each order parameter, characterizing
 a broken symmetry phase, from each of degenerate groundstates
 is shown similar multiple bifurcation behavior.
 Furthermore, to complete the description of an ordered phase,
 it is possible to specify how each order parameter from each of degenerate
 groundstates transforms under a symmetry group that is possessed by the Hamiltonian
 because each order parameter is invariant under only a subgroup of the symmetry
 group
 although the Hamiltonian remains invariant under the full symmetry group.
 Examples are given in the quantum $q$-state Potts models with a
 transverse magnetic field
 by employing the tensor network algorithms based on
 infinite-size lattices.
 For any $q$, a general relation between the local order parameters is found
 to clearly show the subgroup of the $Z_q$ symmetry group.
 In addition, we systematically discuss the criticality in the $q$-state Potts
 model.
\end{abstract}

\pacs{03.67.-a, 05.30.Rt, 05.50.+q, 75.40.Cx}
 %
 %03.67.-a Quantum information:
 %05.30.Rt Quantum phase transitions
 %05.50.+q Lattice theory and statistics (Ising, Potts, etc.)
 %75.40.Cx Static properties (order parameter, static susceptibility, heat capacities, critical exponents, etc.)

\maketitle

\section{Introduction}
 Quantum phase transitions (QPTs) have attracted much attention to
 understand a relationship with quantum information \cite{QPT,Chaikin}.
 Compared to local order parameters
 in the conventional Landau-Ginzburg-Wilson paradigm based
 on the spontaneous symmetry breaking mechanism,
 in connections between quantum information and QPTs,
 quantum entanglement, i.e., a purely quantum correlation being absent in classical
 systems, can be used as an indicator of QPTs
 driven by quantum fluctuations in quantum many-body systems\cite{Landau}.
 Quantum fidelity, based on the basic notions of quantum
 mechanics on quantum measurement, has also provided
 an another way to characterize QPTs\cite{Zhou1,Rams,Mukherjee,Zanardi}.
 In the last few years,
 various quantum fidelity approaches such as fidelity per lattice site (FLS)\cite{Zhou1}, reduced
 fidelity\cite{Liu}, fidelity susceptibility\cite{Gu}, density-functional fidelity\cite{Gu},
 and operator fidelity\cite{Xiao}, have been suggested and implemented
 to explore QPTs.

 Actually,
 the fidelity is a measure of similarity between two quantum states by defining
 a overlap function between them.
 The fact that groundstates in different phases should be orthogonal
 due to their distinguishability in the thermodynamic
 limit allows a fidelity between quantum many-body states in
 different phases signaling QPTs
 because
 an abrupt change of the fidelity is expected across a
 critical point in the thermodynamic limit
 \cite{Zhou1,Rams,Mukherjee,Liu,Gu,Xiao,Zanardi}.
 Thus, the fidelity has great advantages
 to characterize the QPTs in a variety
 of quantum lattice systems because
 the groundstate of a system
 undergoes a drastic change in its structure at a critical point,
 regardless of of what type of internal orders are present in
 quantum many-body states.
 Especially,
 the groundstate
 FLS has been manifested to capture drastic changes of the
 groundstate wave functions around a critical point
 even for those
 cannot be described in the framework of Landau-Ginzburg-Wilson
 theory, such as a Beresinskii-Kosterlitz-Thouless (BKT) transtion and
 topological QPTs in quantum lattice many-body systems \cite{Wang}.

 Even though such latest advances in understanding QPTs have been
 made significantly,
 however, understanding {\it directly} degenerate groundstates originated from
 a spontaneous symmetry breaking as a heart of the
 Landau-Ginzburg-Wilson theory has still been remained in unexplored research regimes.
 Also, recently developed tensor network (TN) algorithms\cite{mps,DMRG,Vidal,Zhou2} in quantum lattice
 systems have made it possible to explore directly degenerate groundstates
 with a randomly chosen initial state subject to
 an imaginary time evolution.
 Based on the tensor network algorithm, indeed,
 Zhao {\it et al.} \cite{Zhao}, for the first time, have detected
 a doubly degenerate groundstate by means of FLS bifurcations
 in quantum Ising model and spin-1/2
 XYX model with transverse magnetic field.
 Also, in various spin lattice models \cite{Bi},
 a doubly degenerate groundstate has been detected.
 Further, in the quantum three-state Potts model
 \cite{Dai}, a three-fold degenerate groundstate has been
 detected by using a bifurcation of FLS and a probability mass distribution
 function.

 When the system has more than three-fold degenerate groundstates,
 the degenerate groundstates are orthogonal one another in the
 thermodynamic limit. A quantum fidelity defined by
 a overlap function between degenerate groundstates may not
 distinguish all the degenerate groundstates properly.
 Then, in this paper,
 we investigate how to detect generally $N$-fold degenerate
 groundstates calculated from a tensor network algorithm.
 To do this, we introduce a quantum fidelity
 between degenerate groundstates and an arbitrary reference state.
 The quantum fidelity corresponds to a projection of each
 groundstate onto the chosen reference state.
 Straightforwardly, for a broken symmetry phase, the number of different projection
 magnitudes denotes the groundstate degeneracy.
 At a critical point, the different projection magnitudes collapse
 to one projection magnitude. In such a property of the defined
 quantum fidelity, the different projection magnitudes
 of the groundstates can be called as multiple bifurcation of the quantum fidelity.
 A multiple bifurcation point is identified to be a critical point.

 As a prototypical example,
 we explore the groundstate wavefunctions in the $q$-state Potts
 model\cite{Potts,Wu} in transverse magnetic fields.
 By employing the infinite matrix product state (iMPS) with the
 time evolving block decimation (iTEBD) method \cite{Vidal},
 we calculate the groundstates of the model.
 Due to the broken $Z_q$ symmetry, the $q$-fold degenerate
 groundstates in the broken symmetry phase are
 obtained by means of the quantum fidelity
 with $q$ branches.
 A continuous (discontinuous) QPT for $q \leq 4$ ($q > 4$)
 has been manifested by a continues (discontinuous) fidelity
 function across the critical
 point.
 The multiple bifurcation points are shown to correspond to
 the critical points.
 Also, we discuss a multiple bifurcation of local order parameters
 and its characteristic properties for the broken symmetry phase.
 We find a general relation between the order parameters from
 each of degenerate groundstates.
 The general relation shows clearly how
 the order parameters from each of the degenerate groundstates
 transform under the subgroup of the
 symmetry group $Z_q$.
 In addition, for $q=4$,
 we calculate the critical exponents
 agree well with their exact values.
 From the finite-entanglement scalings
 of the von Neumann entropy and the correlation length,
 the cental charges are calculated to classify the universal classes
 for each $q$-state Potts model.

 This paper is organized as follows. In Sec.II, we briefly explain the
 iMPS representation and the iTEBD method in one-dimensional quantum lattice systems.
 In Sec. III, the $q$-state Potts model is introduced.
 Section IV devotes how to detect degenerate groundstate
 by using a quantum fidelity between the degenerate groundstates
 and a reference state.
 By using the quantum fidelity per lattice site,
 the quantum phase transitions are discussed based its multiple
 bifurcations and multiple bifurcation points indicating quantum
 critical points in Sec. V.
 In Sec. VI, to complete the description of the ordered phases,
 we discuss the magnetizations given from the degenerate groundstates
 and obtain a general relation between them.
 Section VII presents the critical exponents for $q=4$
 and the central charges for $q=3$ and $q=4$.
 Also, we discuss the quantum phase transitions from the von Neumann
 entropies.
 Finally, our summary is given in Sec. VIII.

\section{Numerical method: IMPS approach}

 Recently, significant progress has been made in
 numerical studies based on TN representations
 for the investigation of quantum phase transitions\cite{mps,DMRG,Vidal,Murg},
 which offers a new perspective from quantum entanglement and fidelity, thus
 providing a deeper understanding on characterizing critical
 phenomena
 in finite and infinite spin lattice systems.
 Actually, a wave function represented in
 TNs allows to perform the classical
 simulation of quantum many-body systems.
 Especially, in one-dimensional spin systems,
 a wave function for infinite-size lattices
 can be described by the iMPS representation\cite{Vidal}.
 which
 has been successfully applied to investigate
 the properties of groundstate wave functions in various infinite spin lattice
 systems\cite{Wang,Liu}.

 For an infinite one-dimensional lattice system,
 a state can be written as\cite{Vidal}
 \begin{eqnarray}
 &&|\Psi\rangle
  =  % \sum_{i}
  \sum_{\{S\}} \sum_{\{\alpha\}}
 \cdots
% \Gamma^{[i-1]}_{\alpha_{i-1},s_{i-1},\alpha_{i}}
 \lambda^{[i]}_{\alpha_i} \Gamma^{[i]}_{\alpha_i,s_i,\alpha_{i+1}}
 \lambda^{[i+1]}_{\alpha_{i+1}} \Gamma^{[i+1]}_{\alpha_{i+1},s_{i+1},\alpha_{i+2}}
 \lambda^{[i+2]}_{\alpha_{i+2}}
\cdots
 \nonumber
 \\
  && ~\hspace*{2.0cm} \times |\cdots S_{i-1}S_{i}S_{i+1}
% s^{[i+2]}s^{[i+3]}
 \cdots \rangle,
 \label{wave}
 \end{eqnarray}
 where $|S_i\rangle$ denote a basis of the local Hilbert space at
 the site $i$, the elements of a diagonal matrix $\lambda^{[i]}_{\alpha_i}$
 are the Schmidt decomposition coefficients of the bipartition
 between the semi-infinite chains $L(-\infty,...,i)$ and
 $R(i+1,...,\infty)$,
 and $\Gamma^{[i]}_{\alpha_i,S_i,\alpha_{i+1}}$ are a three-index
 tensor. The physical indices $S_i$ take the value $1, \cdots, d$
 with the local Hilbert space dimension $d$ at the site $i$.
 The bond indices $\alpha_i$ take the value $1, \cdots, \chi$
 with the truncation dimension of the local Hilbert space at the
 site $i$. The bond indices connect the tensors $\Gamma$ in the nearest neighbor sites.
 Such a representation in Eq. (\ref{wave}) is called
 the iMPS representation.
 If a system Hamiltonian has a translational invariance, one can
 introduce a translational invariant iMPS representation for a state.
 Practically, for instance, for a two-site translational invariance,
 the state can be reexpressed in terms of only the three-index tensors
 $\Gamma_{A(B)}$ and the two diagonal matrices $\lambda_{A(B)}$ for
 the even (odd) sites, where $\{\Gamma, \lambda\}$ are in the canonical form,
 i.e.,
 \begin{eqnarray}
 |\Psi\rangle
  =  % \sum_{i}
  \sum_{\{S\}} \sum_{\{l,r\}}
 \cdots
% \Gamma^{[i-1]}_{\alpha_{i-1},s_{i-1},\alpha_{i}}
 \lambda_A \Gamma_A
 \lambda_B \Gamma_B
 \lambda_A
 \cdots
% \nonumber
% \\
%  && ~\hspace*{2.0cm} \times
  |\cdots S_{i-1}S_{i}S_{i+1}
% s^{[i+2]}s^{[i+3]}
 \cdots \rangle,
 \label{state2}
 \end{eqnarray}
 where $l$ and $r$ are the left and right bond indices, respectively.

 Once a random initial state $|\Psi(0)\rangle$ is prepared in the iMPS representation,
 one may employ the iTEBD
 algorithm to calculate a groundstate wavefunction numerically.
 For instance, if a system Hamiltonian is
 translational invariant and the interaction between spins consists of
 the nearest-neighbor interactions, i.e.,
 the Hamiltonian can be expressed by $H=\sum_{i}h^{[i,i+1]}$, where $h^{[i,i+1]}$
 is the nearest-neighbor two-body Hamiltonian density,
 a groundstate  wavefunction of the system can be expressed in the
 form in Eq. (\ref{state2}).
 The imaginary
 time evolution of the prepared initial state $|\Psi(0)\rangle$, i.e.,
\begin{equation}
|\Psi(\tau)\rangle=\frac{\exp[-H\tau]|\Psi(0)\rangle}{||\exp[-H\tau]|\Psi(0)\rangle||},
\end{equation}
 leads to a groundstate of the system for a large enough $\tau$.
 By using the Suzuki-Trotter decomposition\cite{suzuki},
 actually, the imaginary time evolution
 operator $U=\exp[-H\tau]$
 can be reduced to a product of two-site evolution operators $U(i,i+1)$ that only
 acts on two successive sites $i$ and $i+1$.
 For the numerical imaginary time evolution operation,
 the continuous time evolution can be approximately realized by
 a sequence of the time slice evolution gates $U(i,i+1)=\exp\left[-h^{[i,i+1]}
 \delta\tau\right]$
 for the imaginary time slice $\delta \tau = \tau/n \ll 1$.
 A time-slice evolution gate operation
 contracts $\Gamma_A$, $\Gamma_B$, one $\lambda_A$, two $\lambda_B$,
 and the evolution operator $U(i,i+1)=\exp\left[-h^{[i,i+1]}
 \delta\tau\right]$.
 In order to recover the evolved state in the iMPS representation,
 a singular value decomposition (SVD) is performed
 and the $\chi$ largest singular values are obtained.
 From the SVD,
 the new tensors $\Gamma_A$, $\Gamma_B$, and $\lambda_A$
 are generated. The latter is used to update the tensors $\lambda_A$
 as the new one for all other sites.
 Similar contraction on the new tensors
 $\Gamma_A$, $\Gamma_B$, two new $\lambda_A$, one  $\lambda_B$,
 and the evolution operator $U(i+1,i+2)=\exp\left[-h^{[i+1,i+2]}\right]$,
 and its SVD produce the updated $\Gamma_A$, $\Gamma_B$, and $\lambda_B$
 for all other sites.
 After the time-slice evolution, then,
 all the tensors $\Gamma_A$, $\Gamma_B$, $\lambda_A$, and $\lambda_B$
 are updated.
 This procedure is repeatedly performed until the system energy
 converges to a groundstate energy that yields a groundstate
 wavefunction in the iMPS representation.
 The normalization of the groundstate wavefunction is guaranteed by
 requiring the norm $\langle \Psi | \Psi \rangle = 1$.
 Finally, one can determine how many groundstates exist for a fixed
 system parameters with different initial states.
 Once a system undergoes a spontaneous symmetry breaking,
 the iMPS algorithm can automatically produce degenerate groundstates
 with randomly chosen initial states for the broken symmetry phase.
 It has been manifested by successfully detecting a doubly degenerate
 groundstates from $Z_2$ broken symmetry phases in various spin systems
 such as quantum Ising model, spin-1/2 XYX with transverse magnetic field,
 and a one-dimensional spin model with competing
 two-spin and three-spin interactions \cite{Zhao,Bi}.
 However,
 it has not been discussed yet how to detect an $N$-fold degenerate groundstates.
 In the following sections,
 we discuss this issue clearly by introducing the $q$-state Potts
 model that may have a $q$-fold generate groundstate in broken symmetry phases.

 \section{Quantum $q$-state Potts model}
 In the lattice statistical mechanics,
 the quantum $q$-state Potts model is
 one of the most important models
 as a generalization of the Ising model ($q=2$)\cite{Baxter2,Wu}.
 The $q$-state Potts model has the very intriguing
 critical behavior that has become an
 important testing platform for different numerical and analytical
 methods and approaches in
 studying critical phenomena\cite{Wu,Nijs,Pearson,Black,Caselle,Hove}.
 As is well-known in Ref. \onlinecite{Wu},
 the quantum q-state Potts model exhibits a continuous quantum phase
 transition for $q \leq 4$ and a first-order (discontinuous) phase transition
 for $q > 4$ at the critical point.

 We consider the $q$-state quantum Potts model in a
 transverse magnetic field $\lambda$ on an infinite-size lattice:
 \begin{equation}
 H_q=-\sum_{j=1}^{\infty}\sum^{q-1}_{p=1}\Bigg(M_{x,\, p}(j)~M_{x,\, q-p}(j+1)+\lambda
 M_{z}(j)\Bigg),
 \label{ham}
 \end{equation}
 where $\lambda$ is the transverse magnetic field and
 $M_{x/z,p}(i)$ with $p \in [ 1, q-1] $
 are the $q$-state Potts spin matrices at the lattice site $j$.
 The $q$-state Potts spin matrices are
 given as
 \begin{equation}\nonumber
  M_{x,1}= \left(\begin{array}{ccccc}
                  0 & I_{q-1} \\ 1 & 0
 \end{array} \right)
 \mbox{~and~}
  M_{z}= \left(\begin{array}{ccccc}
                  q-1 & 0 \\ 0 & I_{q-1}
 \end{array} \right),
 \end{equation}
 where $I_{q-1}$ is the $(q-1) \times (q-1)$ identity matrix and
 $M_{x,p}=(M_{x,1})^p$.
 As is known, the model Hamiltonian has a $Z_q$ symmetry\cite{Wu}.
 If the groundstate of the Hamiltonian does not preserve the $Z_q$
 symmetry, the system undergoes the $Z_q$ symmetry breaking, i.e.,
 a QPT.
 Specifically, when the magnetic filed changes across $\lambda_c =
 1$, the $q$-state Potts model undergoes a QPT between an ordered phase
 and a disordered phase.
 The QPT is originated from the $Z_q$ symmetry breaking which
 results in the emergence of long-range order and
 the $q$ degenerate groundstates in the ordered phase.

 \section{Degenerate groundstates and quantum fidelity}

%%%%%%%%%%%%%%%%%%%%%%%%%%%%%%%%%%%%%%%%%%%%%%%%%%%%%%%%%%%%%%
 \begin{figure}
 \begin{center}
 \includegraphics[width=3.in]{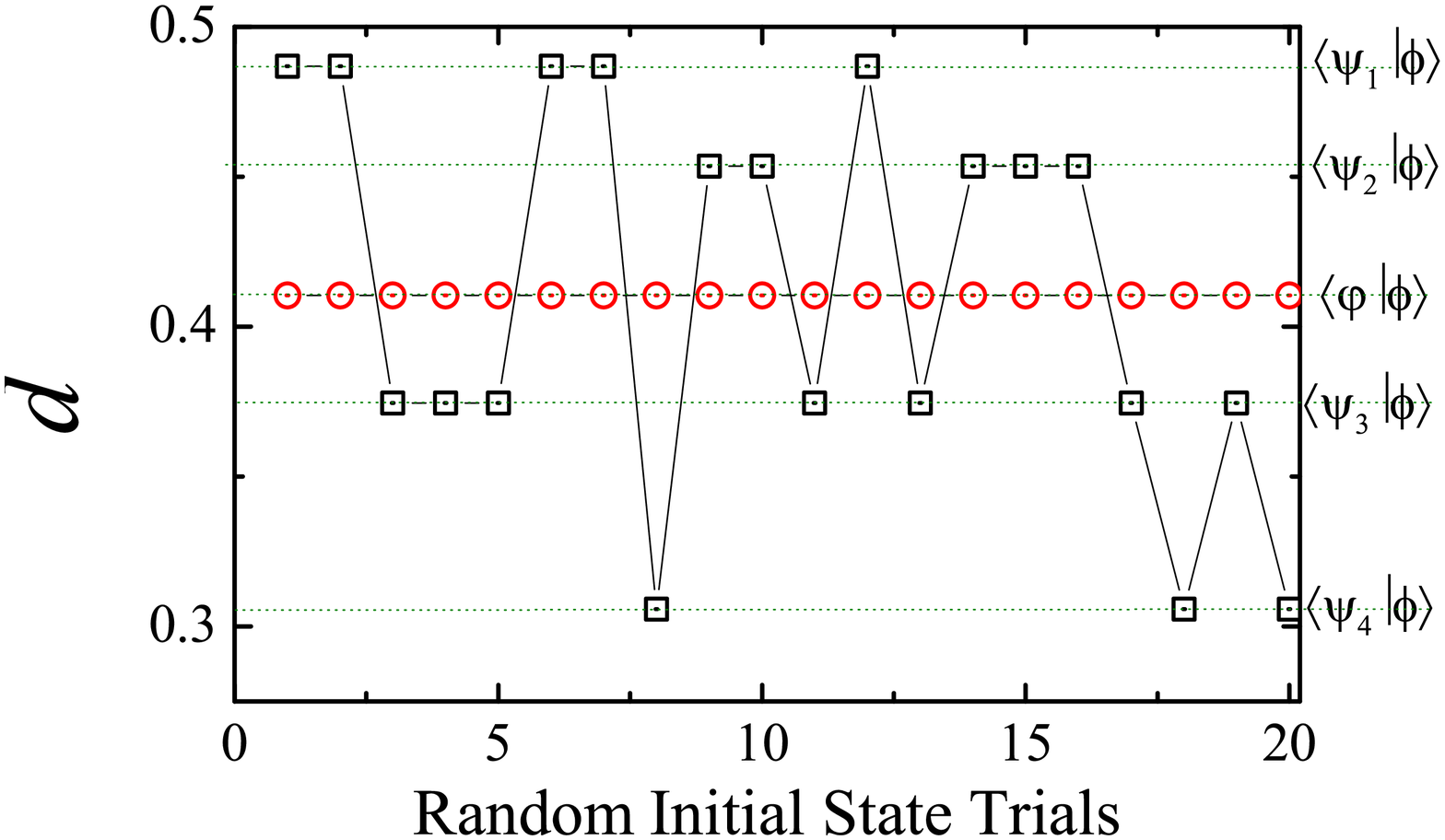}
 \end{center}
 \caption{(Color online) Groundstate quantum FLS
 $d$ for the $4$-state Potts model with an arbitrary reference state as a
 function of random initial state trials.
 Here, $|\phi\rangle$ is an arbitrary reference state,
 the numerical groundstates $|\psi\rangle$
 are in the broken symmetry phase with $\lambda=0.8$, and $|\varphi\rangle$
 is in the symmetry phase with $\lambda=1.2$, respectively.
 It is clearly shown that,
 there exist four degenerate groundstates (black rectangle)
 which are labeled
 by $|\psi_1\rangle$, $|\psi_2\rangle$, $|\psi_3\rangle$, and
 $|\psi_4\rangle$.
 In the broken symmetry
 phase for $\lambda=1.2$, only one groundstate (red circle) exists.
 The dotted lines are guided the eyes. If
 the number of random initial state trials increases,
 the probability that the system is in each degenerate groundstate
 approaches $1/4$ in the broken symmetry
 phase for the $4$-state Potts model.
 }\label{fig1}
 \end{figure}
%%%%%%%%%%%%%%%%%%%%%%%%%%%%%%%%%%%%%%%%%%%%%%%%%%%%%%%%%%%%%%%%%%%%%%%%

%
 From a tensor network approach with an infinite lattice system,
 once one obtains a groundstate $|\psi^{(n)}\rangle$
 with the $n$-th random initial state, one can define a quantum fidelity
  $F(|\psi^{(n)}\rangle,|\phi\rangle)=|\langle\psi^{(n)}|\phi\rangle|$
  between the groundstate and a chosen reference state $|\phi\rangle$.
 If $F(|\psi^{(n)}\rangle,|\phi\rangle)$ has only one constant value with random
 initial states, the system has only one groundstate.
 If $F(|\psi^{(n)}\rangle,|\phi\rangle)$ has $N$ projection values with random
 initial states, the system must have $N$ degenerate groundstates.
 To distinguish the degenerate groundstates,
 we employ the groundstate FLS in
 following Ref. \onlinecite{Zhou1} as
\begin{equation}
 \ln d(|\psi^{(n)}\rangle,|\phi\rangle) \equiv \lim_{L \rightarrow \infty}
  \frac{\ln F(|\psi^{(n)}\rangle,|\phi\rangle)}{L},
 \label{fidelity}
\end{equation}
 where $L$ is the system size.
 The FLS is well defined in the thermodynamic limit even if
 $F$ becomes trivially zero.
 From the fidelity $F$,
 the FLS has several properties as
 (i) $d(|\psi^{(n)}\rangle=|\phi\rangle)=1$  and (ii) its range $0 \le
 d(|\psi^{(n)}\rangle,|\phi\rangle)\le 1$.
 Within the iMPS approach,
 the FLS $d(|\psi^{(n)}\rangle,|\phi\rangle)$ is given by the largest eigenvalue $\mu_0$ of
 the transfer matrix $T$ up to the corrections that decay exponentially
 in the linear system size $L$. Then, for the infinite-size system,
 $d(|\psi^{(n)}\rangle,|\phi\rangle)= \mu^{(n)}_0$.
 Actually, in order to study quantum critical phenomena in quantum lattice systems,
 Zhou and Barjaktarevi\v{c} defined the FLS
 from the fidelity between groundstates \cite{Zhou1,Zhou2}.
 The Zhou-Barjaktarevi\v{c} FLS
  has been manifested as a model-independent and universal
 indicator successfully detecting quantum phase
 transition points including KT transition and topological phase
 transition\cite{Wang}.
 The FLS introduced in Eq. (\ref{fidelity}) for this
 study is a simple extension of the Zhou-Barjaktarevi\v{c} FLS.
 By means of a bifurcation of the Zhou-Barjaktarevi\v{c} FLS
 and a probability mass distribution function,
 one can distinguish degenerate groundstates and detect a quantum phase transition.

%%%%%%%%%%%%%%%%%%%%%%%%%%%%%%%%%%%%%%%%%%%%%%%%%%%%%%%%%%%%%%%%%%%%%%%%%%%%%%
 \begin{figure}
 \begin{center}
 \includegraphics[width=3.in]{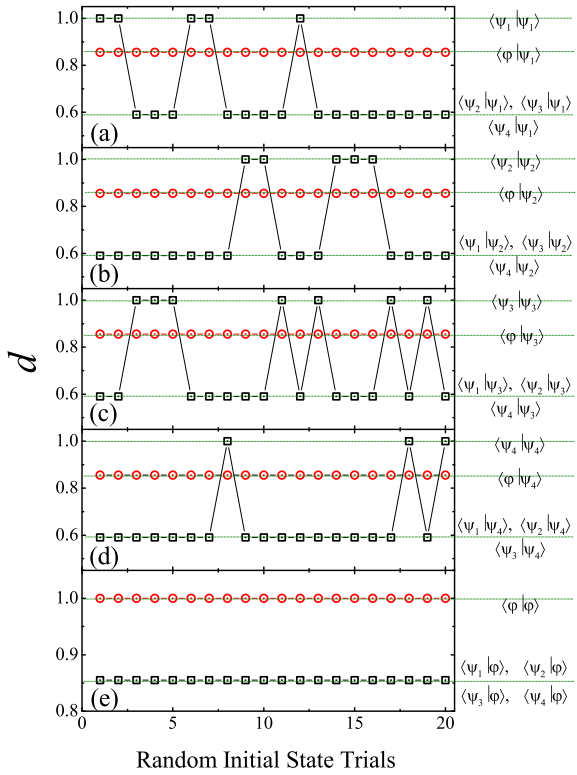}
 \end{center}
 \caption{(Color online) Groundstate quantum FLS
 $d$ for the $4$-state Potts model as a
 function of random initial state trials.
 The reference states are chosen in (a) $|\Psi_1\rangle$,
 (b) $|\Psi_2\rangle$, (c) $|\Psi_3\rangle$,
 (d) $|\Psi_4\rangle$, and (e) $|\varphi\rangle$.
 It is clearly shown that, if the reference state is chosen as
 one of the degenerate groundstates in the broken symmetry phase,
 the other three groundstates has the same value of the FLS in (a)-(d).
 If the reference state is chosen as
 the groundstate in the symmetry
 phase, the FLSs for
 the four degenerate groundstates has the same value in the broken symmetry
 phase in (e).
 }\label{fig2}
 \end{figure}
%%%%%%%%%%%%%%%%%%%%%%%%%%%%%%%%%%%%%%%%%%%%%%%%%%%%%%%%%%%%%%%%%%%%%%%%%%%%%%%%

 In order to show clearly how to detect degenerate groundstates based on Eq.
 (\ref{fidelity}),
 as an example, we consider the $4$-state Potts model.
 In Fig. \ref{fig1}, we plot the FLS $d$ for random initial
 states. Here, $\lambda = 0.8$ and $1.2$ are chosen.
 Due to the broken $Z_4$ symmetry for $q=4$,
 the system has $4$ degenerate groundstats for the broken symmetry
 phase $\lambda < \lambda_c=1$.
 It is clearly shown that, for $\lambda=0.8$ $(\lambda < \lambda_c)$,
 there exist four different values of
 the FLS, while, for $\lambda=1.2$ $(\lambda > \lambda_c)$,
 there exists only one value of the FLS.
 From each value of the FLS for $\lambda=0.8$,
 we label the four degenerate groundstates
 as $|\psi_1\rangle$, $|\psi_2\rangle$, $|\psi_3\rangle$, and
 $|\psi_4\rangle$. For for $\lambda=1.2$, the groundstate is denoted
 by $|\varphi\rangle$.
 Actually, we have chosen more than 200 random initial states.
 The probability $P(n)$, that the system is in each groundstate
 for  $\lambda < \lambda_c$, is found to be
 $P(n) \simeq 1/4$ in the broken symmetry phase.
 Then, for the $q$-state Potts model,
 with a large number of random initial state trials,
 one may detect the $q$ degenerate groundstates
 with the probability $P(n \rightarrow \infty)=1/q$ finding each degenerate groundstate
 in the broken $Z_q$ symmetry phase.

 One may choose the reference state as one of the degenerate
 groundstates.
 In Fig. \ref{fig2},
 we choose the reference state as one of the groundstates
 for the broken symmetry phase in (a)
 $|\Psi_1\rangle$, (b) $|\Psi_2\rangle$,
 (c) $|\Psi_3\rangle$, and (d) $|\Psi_4\rangle$, and
 as the groundstate  $|\varphi\rangle$ for the symmetry phase in (e).
  In Figs. \ref{fig2}(a)-(d),
  it is clearly shown that, for $\lambda=0.8$ $(\lambda < \lambda_c)$,
 there exist two different values of
 the FLS, i.e., $d(|\psi_m\rangle,|\psi_m\rangle)=1$ and
 $d(|\psi_m\rangle,|\psi_{m\neq m'}\rangle)=constant \neq 1$
 $(m, m' \in [1,q] )$.
 Furthermore, $d(|\psi_m\rangle,|\varphi\rangle)=constant$.
 In Fig. \ref{fig2} (e),
 the reference state chosen as the groundstate
 in the symmetry phase shows
 $d(|\varphi\rangle,|\varphi\rangle)=1$ and
 $d(|\psi_m\rangle,|\varphi\rangle)=constant \neq 1$.
 In this case, as discussed in Ref. \onlinecite{Dai},
 one may use a probability mass distribution function,
 as an alternative way, to distinguish the degenerate groundstates.
 Here, it is shown that the degenerate states can be distinguished by
 a reference state chosen as arbitrary states except for
 the degenerate groundstates in the broken symmetry phase
 and the groundstate in the symmetry phase.

%%%%%%%%%%%%%%%%%%%%%%%%%%%%%%%%%%%%%%%%%%%%%%%%%%%%%%%
 \begin{figure}
 \begin{center}
 \includegraphics[width=3.in]{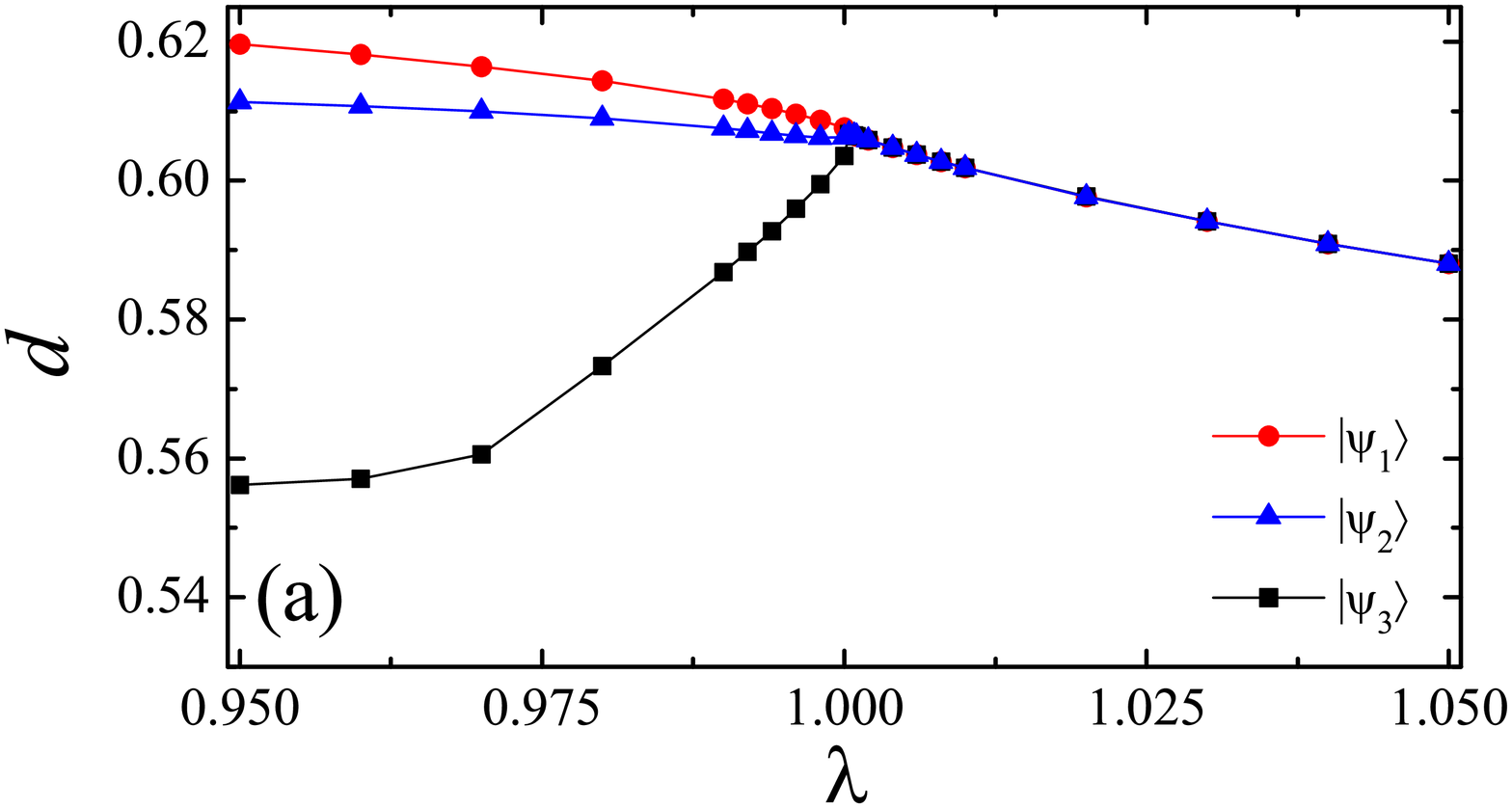}
 \includegraphics[width=3.in]{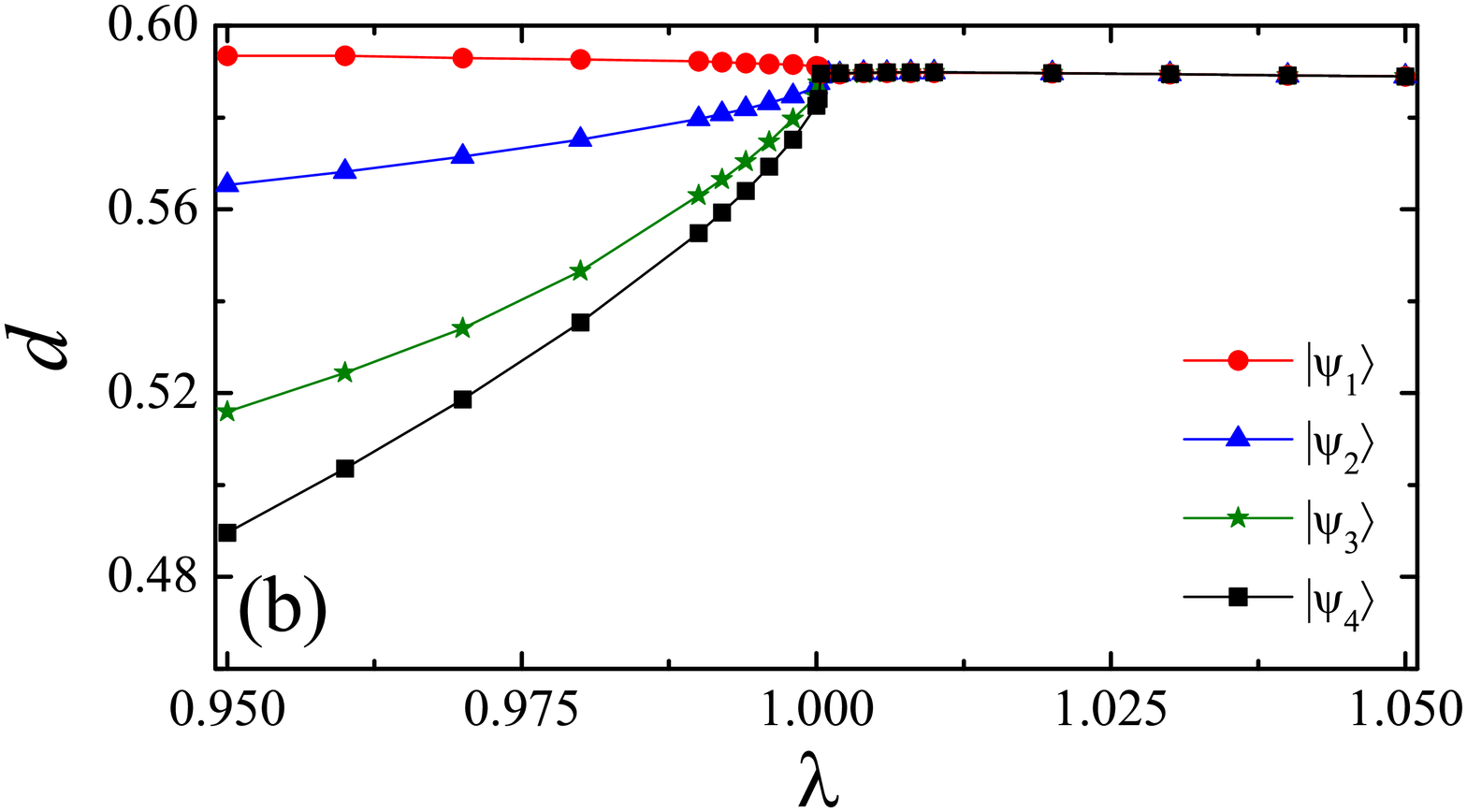}
  \includegraphics[width=3.in]{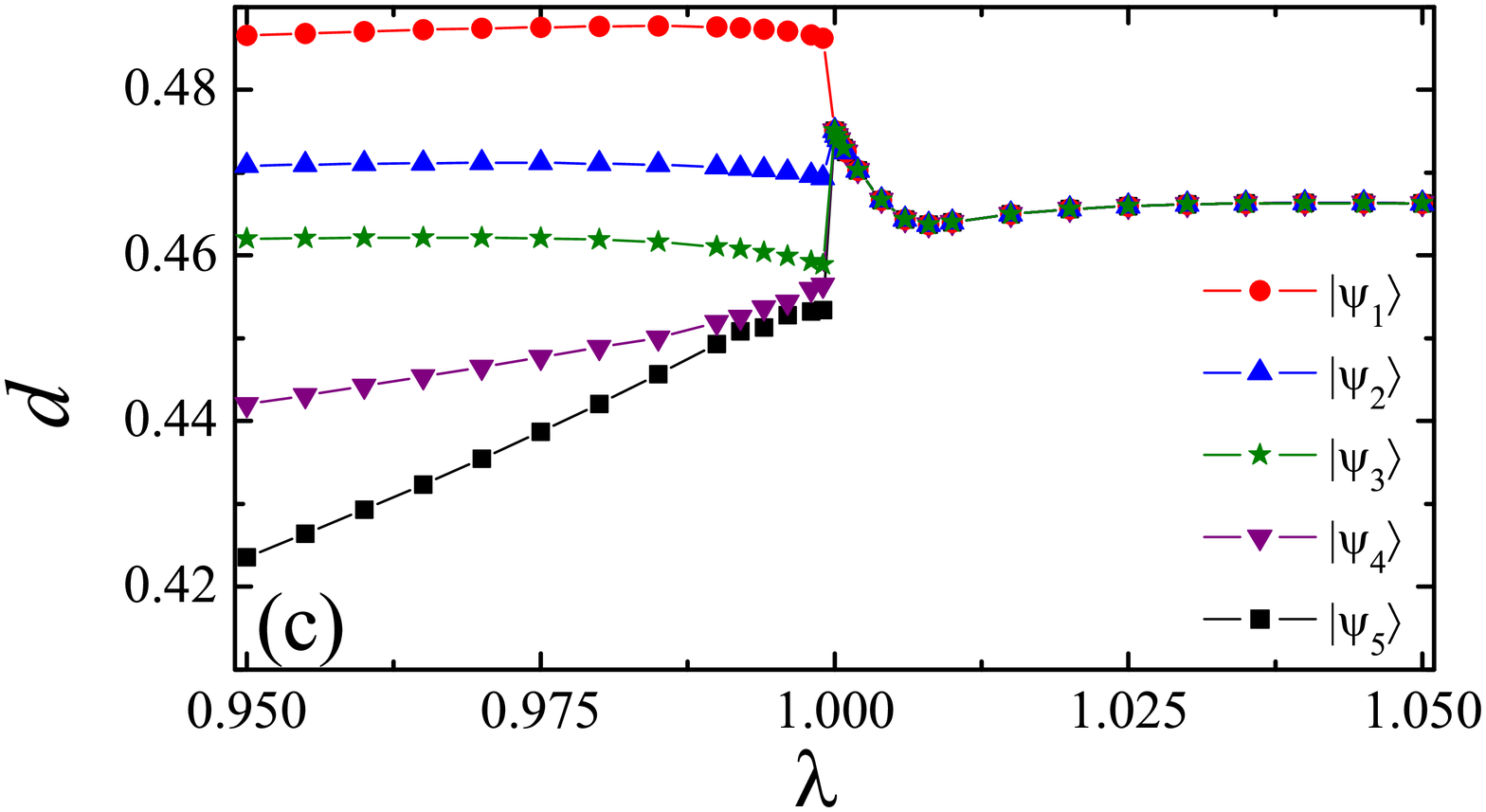}
 \end{center}
 \caption{(Color online) Groundstate quantum FLS
 $d$ for (a) $3$-, (b) $4$-, and (c) $5$-state Potts models
 as a function of the transverse magnetic field
 $\lambda$ with the truncation dimension $\chi =32$.
 In the broken symmetry phase,
 the $q$ branches of the FLS correspond to the numbers of the $q$ degenerate groundstates.
 As the magnetic field crosses the critical point $\lambda_c$,
 the FLSs show the multiple bifurcations with three,
 four, and five branches, respectively, in the broken symmetry phase.
 Note that, for the $3$- and $4$-state Potts models,
 the FLSs
 show their continues behaviors, while, for the $5$-state
 Potts model, the FLS
 shows a discontinues behavior. These continues and discontinues
 behaviors indicate the continues phase transitions for $q=3$ and $4$,
 and the discontinues phase transition for $q=5$.
 The numerical multiple bifurcations are given as $\lambda_c=1.0004$ for
 $q=3$ and $4$ and $\lambda_c =1$ for $q=5$.
 }\label{fig3}
 \end{figure}
%%%%%%%%%%%%%%%%%%%%%%%%%%%%%%%%%%%%%%%%%%%%%%%%%%%%%%%%%%%%%%%%%%%%%%

\section{Quantum Fidelity per lattice site for phase transitions}

 In the view of a groundstate degeneracy,
 the degenerate groundstates in the broken symmetry phase
 exist until the system reaches its critical point.
 Once one can detect degenerate groundstates,
 one may also detect a quantum phase transition by the quantum fidelity
 in Eq. (\ref{fidelity}).
 Detected degenerate groundstate wavefunctions
 may allow us also to investigate directly a property of quantum phases
 even for unexplored broken symmetry phases.
 In this section, then, we will discuss the quantum phase transitions
 for the $q$-state Potts model. In the next section, we will discuss
 the relation between system symmetry and order parameter directly
 from groundstates.

 From the detected degenerate groundstates,
 in Fig. \ref{fig3},
 we plot the FLSs as a function of the transverse magnetic field
 $\lambda$ for $q=3$, $4$, and $5$.
 Here, the truncation dimension is chosen as
 $\chi =32$.
 Figure \ref{fig3} shows clearly that, as the transverse magnetic field decreases,
 the FLSs in the symmetry phase
 branch off the $q$ FLSs.
 The branch points are estimated numerically as $\lambda_c=1.0004$
 for the $3$- and $4$-state Potts models.
 Also, for the
 $5$-state Potts model, the multiple bifurcation point
 exists exactly at $\lambda_c =1$.
 These results show that the branch points agree well
 the exact critical point $\lambda_c=1$.
 Then,
 the branch points correspond to the critical point.
 Such a branching behavior of the FLS can be called as {\it multiple
 bifurcation} and a branch point as {\it multiple bifurcation
 point}.
 Consequently, it is shown that
 the FLS from the quantum fidelity between degenerate groundstates
 and a reference state
 can detect a quantum phase transition.

 As is known,
 for the $q$-state Potts model,
 the continues (discontinuous) phase transitions
 occurs for $q \leq 4$ ($q > 4$).
 Figs. \ref{fig3} (a) for $q=3$ and (b) $q=4$ show that
 the FLSs are a continues function for the quantum phase transition.
 While Fig. \ref{fig3} (c) $q=5$ shows that
 the FLS is a discontinues function for the quantum phase transition.
 Then, the continues (discontinues) behaviors at the critical points
 imply that a continues (discontinues) phase transition occurs.
 As a result, the FLS in Eq. (\ref{fidelity}) can clarify continues (discontinues) quantum
 phase transitions.

\section{Symmetry and order parameter}

%%%%%%%%%%%%%%%%%%%%%%%%%%%%%%%%%%%%%
  \begin{figure}
 \begin{center}
 \vspace*{-0.5cm}
 \includegraphics[width=2.5in]{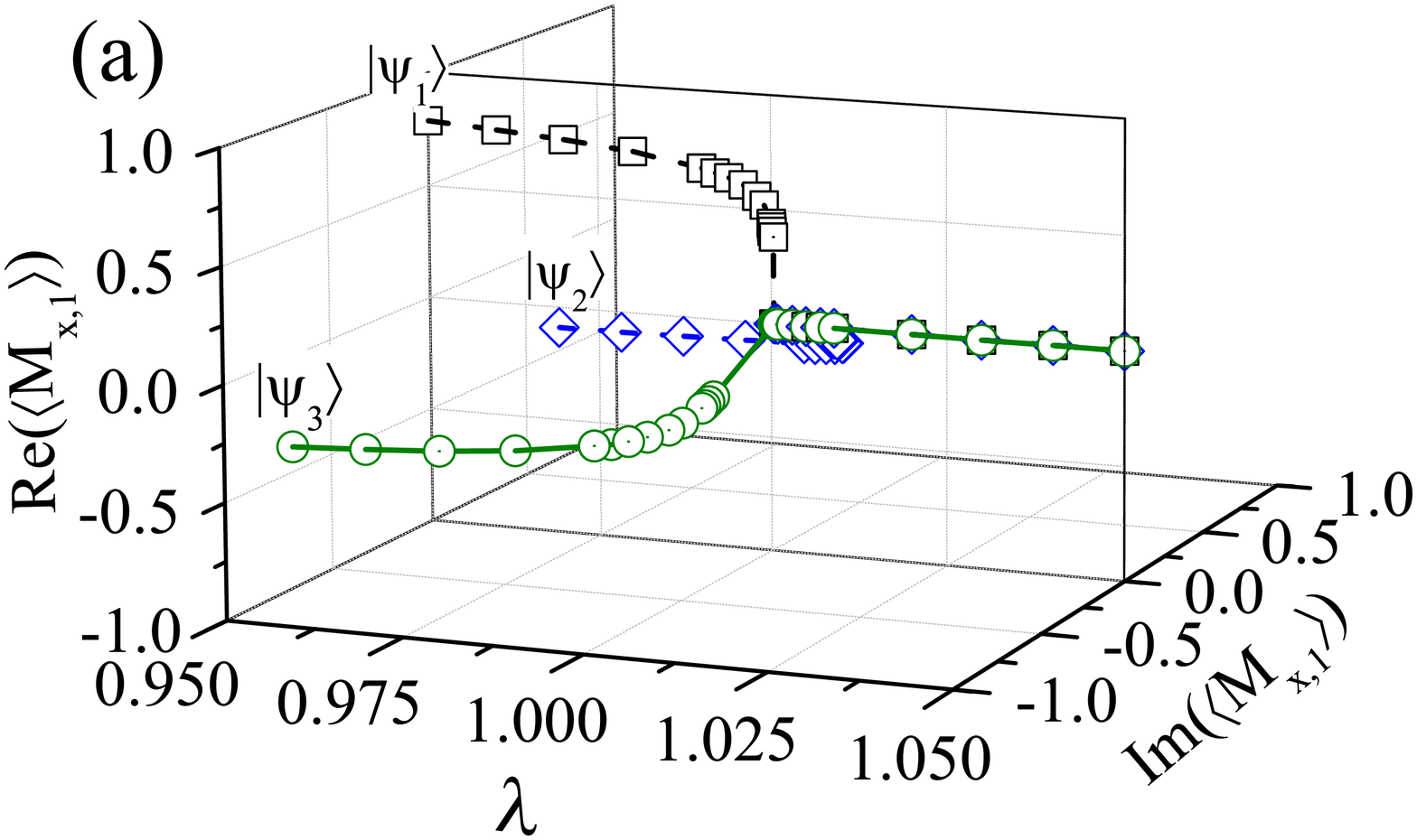}
 \vspace*{-0.5cm}
 \includegraphics[width=2.5in]{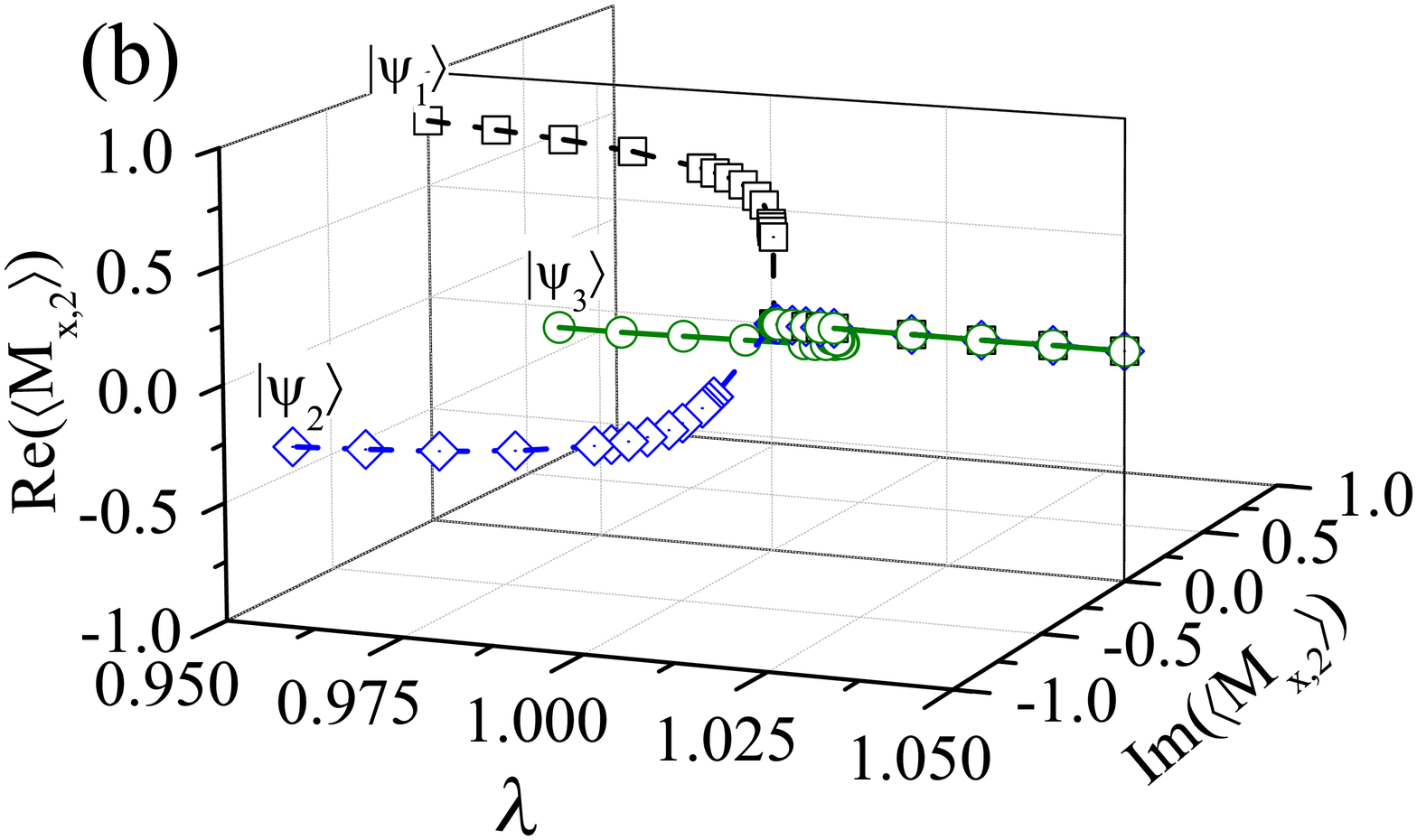}
 \end{center}
 \caption{(Color online) Magnetization (a) $\langle M_{x,1}
 \rangle$ and (b) $\langle M_{x,2} \rangle$ as a function of the transverse magnetic field
 $\lambda$ for the 3-state Potts model.
 For the broken symmetry phase $\lambda < \lambda_c$,
 each magnetization is given for each of the three degenerate groundstates
 denoted by $|\psi_m\rangle$ $(m \in [1,q]$.
 The numerical critical point locates at $\lambda_c=1.0004$.
 }\label{fig4}
 \end{figure}
%%%%%%%%%%%%%%%%%%%

 The existence of a degenerate groundstate means
 that each of degenerate groundstates possess its own order described
 by each corresponding order parameter.
 The each order parameter, which is nonzero value only in an ordered phase,
 should distinguish an ordered phase from
 disordered phases\cite{Michel}.
 To complete the description of an ordered phase,
 it is required to specify how each order parameter from each of degenerate
 groundstates transforms under a symmetry group $G$ that is possessed by the Hamiltonian
 because each order parameter is invariant under only a subgroup of the symmetry
 group $G$
 although the Hamiltonian remains invariant under the full symmetry group $G$
 \ \cite{Chaikin}.
 Then, in this section, we discuss local magnetizations obtained
 from each of the degenerate groundstates.

 Let us first discuss the local magnetizations for the $3$-state Potts
 model, i.e., $q=3$.
 In Fig. \ref{fig4}, we plot the magnetization (a) $\langle M_{x,1}
 \rangle$ and (b) $\langle M_{x,2}\rangle$ as a function of the traverse
 magnetic field $\lambda$.
 The magnetizations disappear to zero at the numerical critical
 point $\lambda_c=1.0004$.
 Also, all the magnetizations show that the phase transition
 is a continuous phase transition.
 For the broken symmetry phase $\lambda < \lambda_c$,
 each magnetization is calculated from each of the three degenerate
 groundstates, which is denoted by $|\psi_m\rangle$.
 Note that all the absolute values of the magnetizations
 $\langle \psi_m | M_{x,p} | \psi_m \rangle \equiv \langle M_{x,p}\rangle_m$
 are the same values at a given magnetic field.
 Furthermore, for a given magnetic field, the magnetizations in the complex
 magnetization plane seem to have a relation between them under a rotation,
 which is characterized by the value $\omega_3 = \exp[2\pi i/3]$.
 Then, in Figs. \ref{fig4}(a) and (b), it is observed that,
 for a given magnetic field $\lambda < \lambda_c$,
 there are the relations between the magnetizations as
 $\langle M_{x,1}\rangle_1=\omega^{-1}_3 \langle M_{x,1}\rangle_2
 =\omega^{-2}_3 \langle M_{x,1}\rangle_3$
 and
 $\langle M_{x,2}\rangle_1=\omega^{-2}_3 \langle M_{x,2}\rangle_2
 =\omega^{-4}_3 \langle M_{x,2}\rangle_3$.
 Also, for each groundstate wavefunction, the magnetizations
 have the relations as
 $\langle M_{x,1}\rangle_1=\langle M_{x,2}\rangle_1$,
 $\langle M_{x,1}\rangle_2=\omega^{-1}_3 \langle M_{x,2}\rangle_2$, and
 $\langle M_{x,1}\rangle_3=\omega^{-2}_3 \langle M_{x,2}\rangle_3$.
 These results show that, in the complex magnetization plane,
 the rotations between the magnetizations
 for a given magnetic field
 are determined by the characteristic rotation angles $\theta = 0$, $2\pi/3$,
 and $4\pi/3$, i.e.,
 $\langle M_{x,p}\rangle_m=g_3 \langle M_{x,p'}\rangle_{m'}$ with
 $g_3 \in \{ I, \omega_3, \omega^2_3 \}$.
 Here, we have chosen the $|\psi_1\rangle$ that gives
 a real value of the magnetizations, i.e.,
 $\langle M_{x,1}\rangle_1$ and $\langle M_{x,2}\rangle_1$ are real.
 Also, the degenerate groundstates give the same values for the
 $z$-component magnetizations, i.e., $\langle M_{z}\rangle_1=\langle M_{z}\rangle_2
 =\langle M_{z}\rangle_3$.

\begin{widetext}
%
%%%%%%%%%%%%%   one coloum  %%%%%%%%%%%%%%%%%%%%%%%%%%%%

%%%%%%%%%%%%%%%%%%%     one coloum  %%%%%%%%%%%%%%%%%%%%%%%%
 \begin{figure}
 \begin{center}
 \vspace*{-0.5cm}
 \includegraphics[width=2.5in]{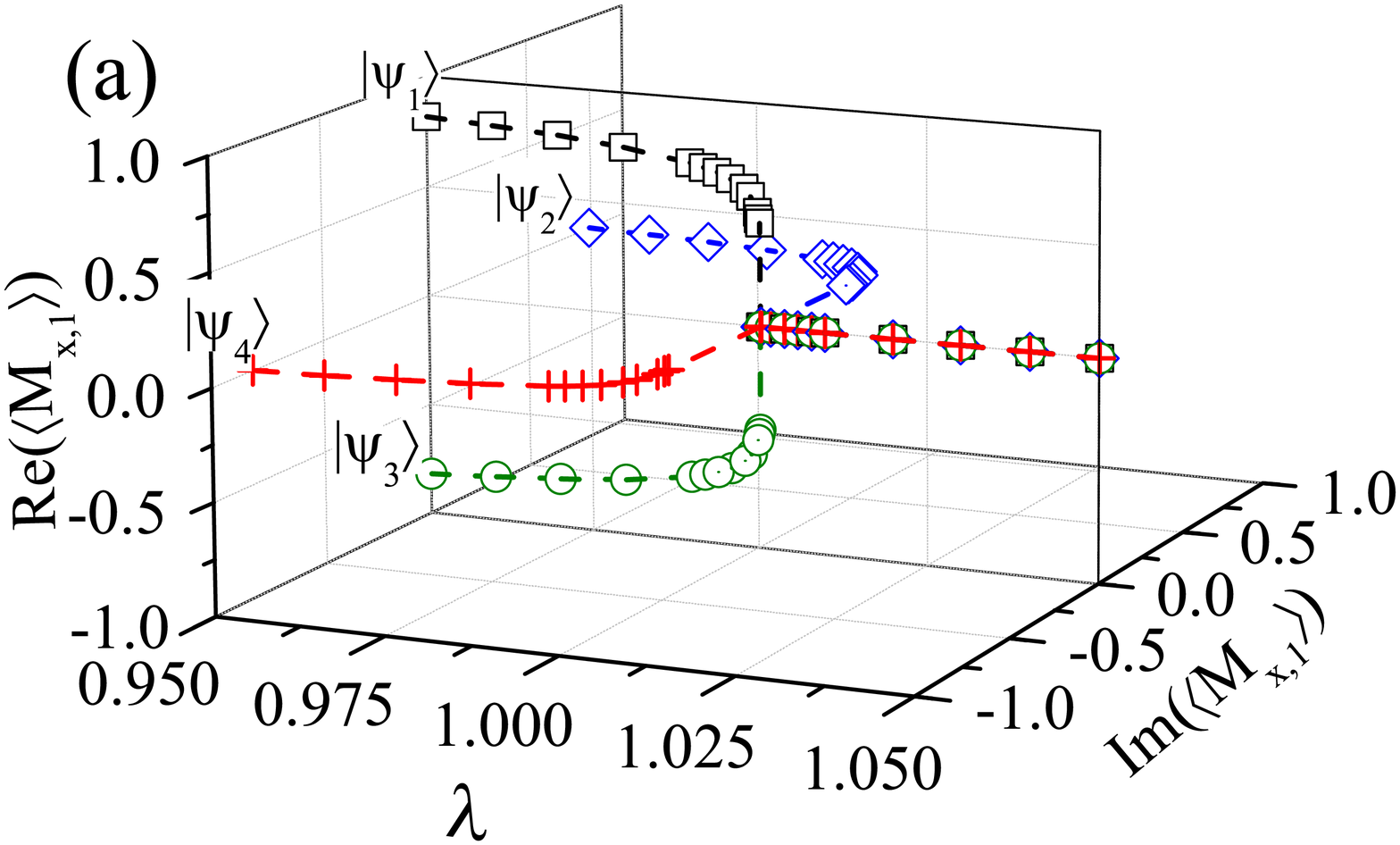}
 \vspace*{-0.5cm}
 \includegraphics[width=2.5in]{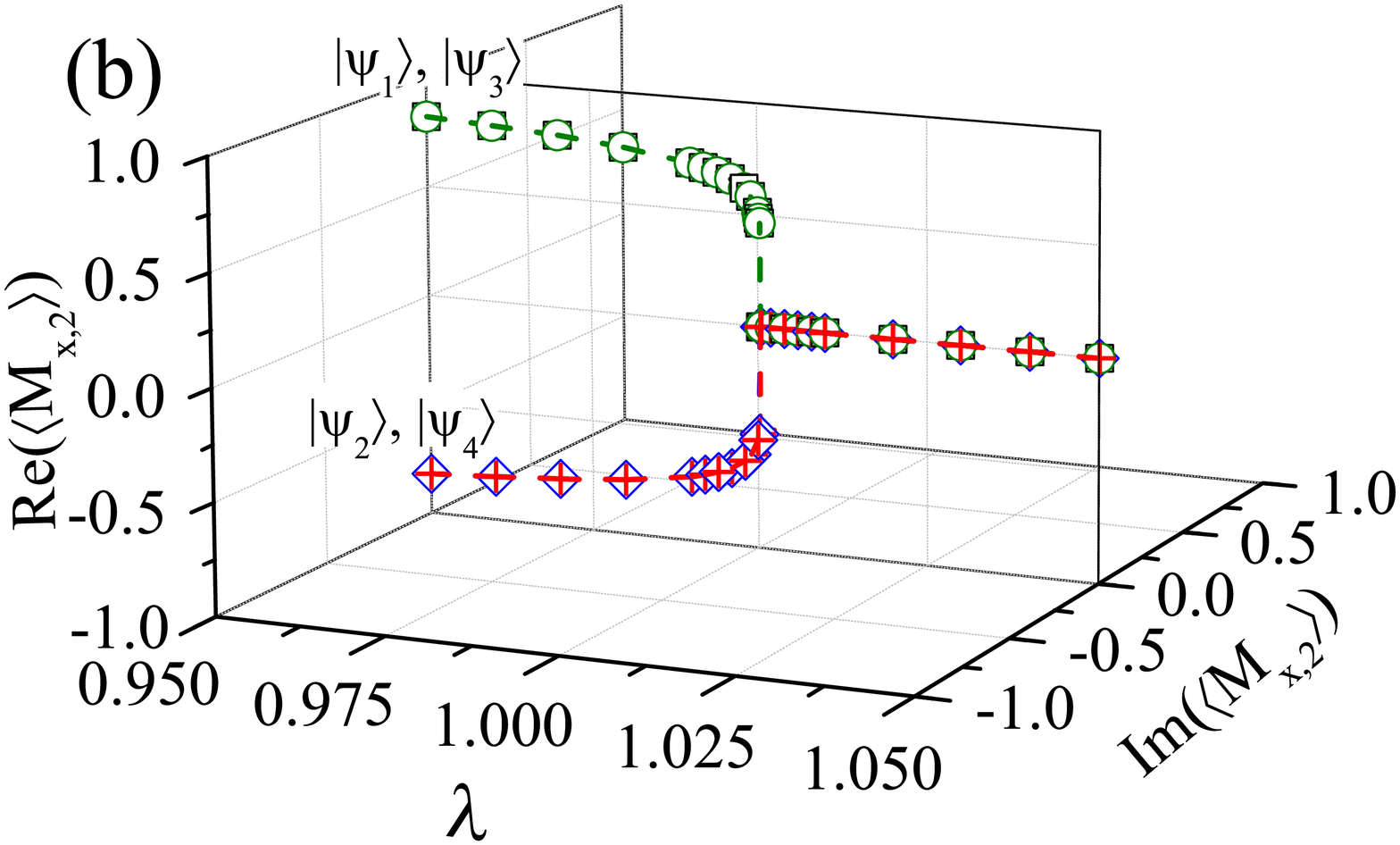}
  \includegraphics[width=2.5in]{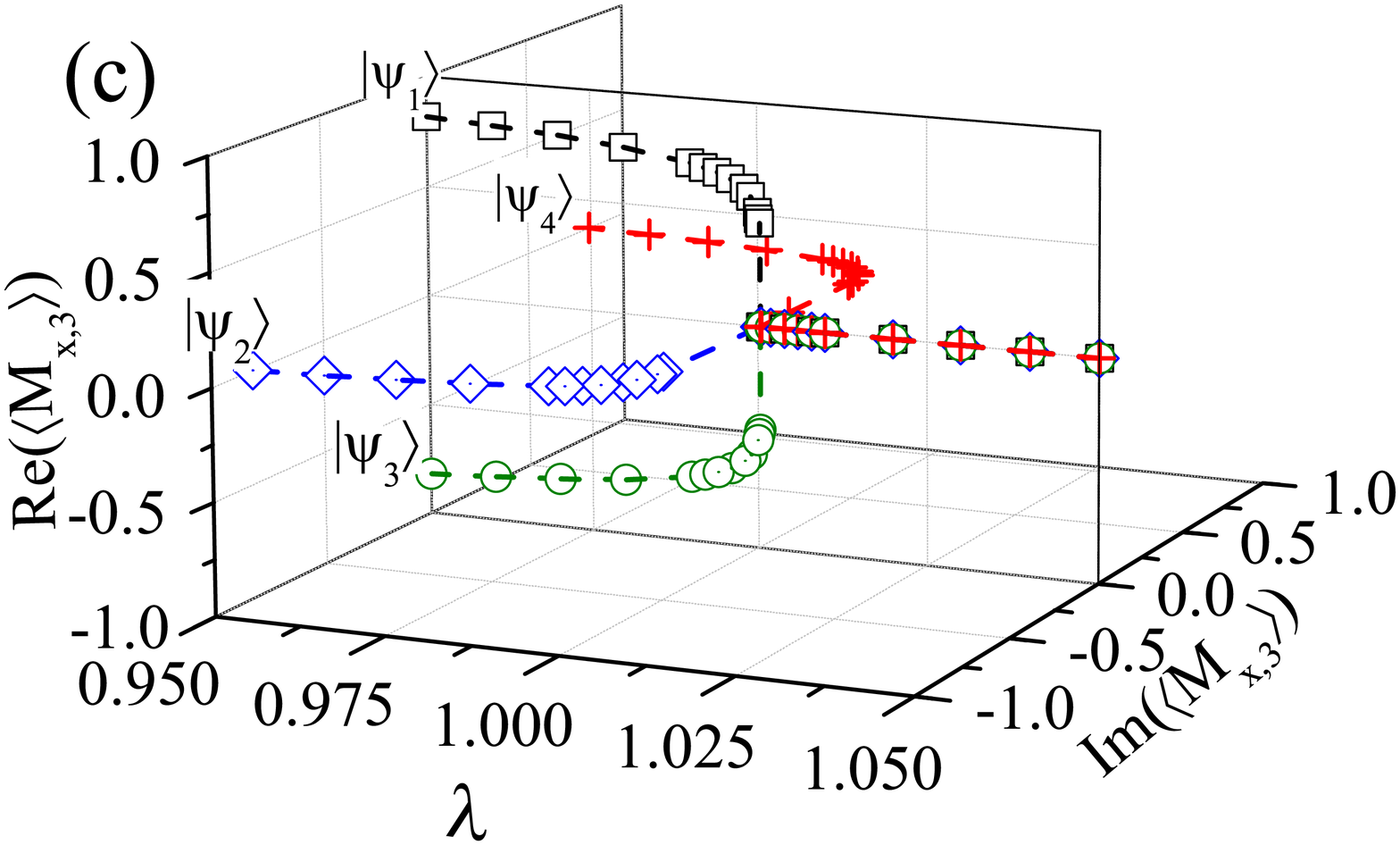}
  \vspace*{-0.5cm}
 \end{center}
  \caption{(Color online) Magnetizations (a) $\langle M_{x,1}
 \rangle$, (b) $\langle M_{x,2} \rangle$,
 and (c) $\langle M_{x,3} \rangle$ as a function of the transverse magnetic field $\lambda$
 for the $4$-state Potts model.
 For the broken symmetry phase,
 the magnetizations are given from each of the four degenerate groundstates.
 The critical point is estimated numerically as $\lambda_c=1.0004$.
  }\label{fig5}
 \end{figure}
%%%%%%%%%%%%%%%%%%%     one coloum  %%%%%%%%%%%%%%%%%%%%%%%%

%%%%%%%%%                one coloum   %%%%%%%%%%%%%%%%%%%%%%%
\begin{figure}
 \begin{center}
 \vspace*{-0.5cm}
  \includegraphics[width=2.5in]{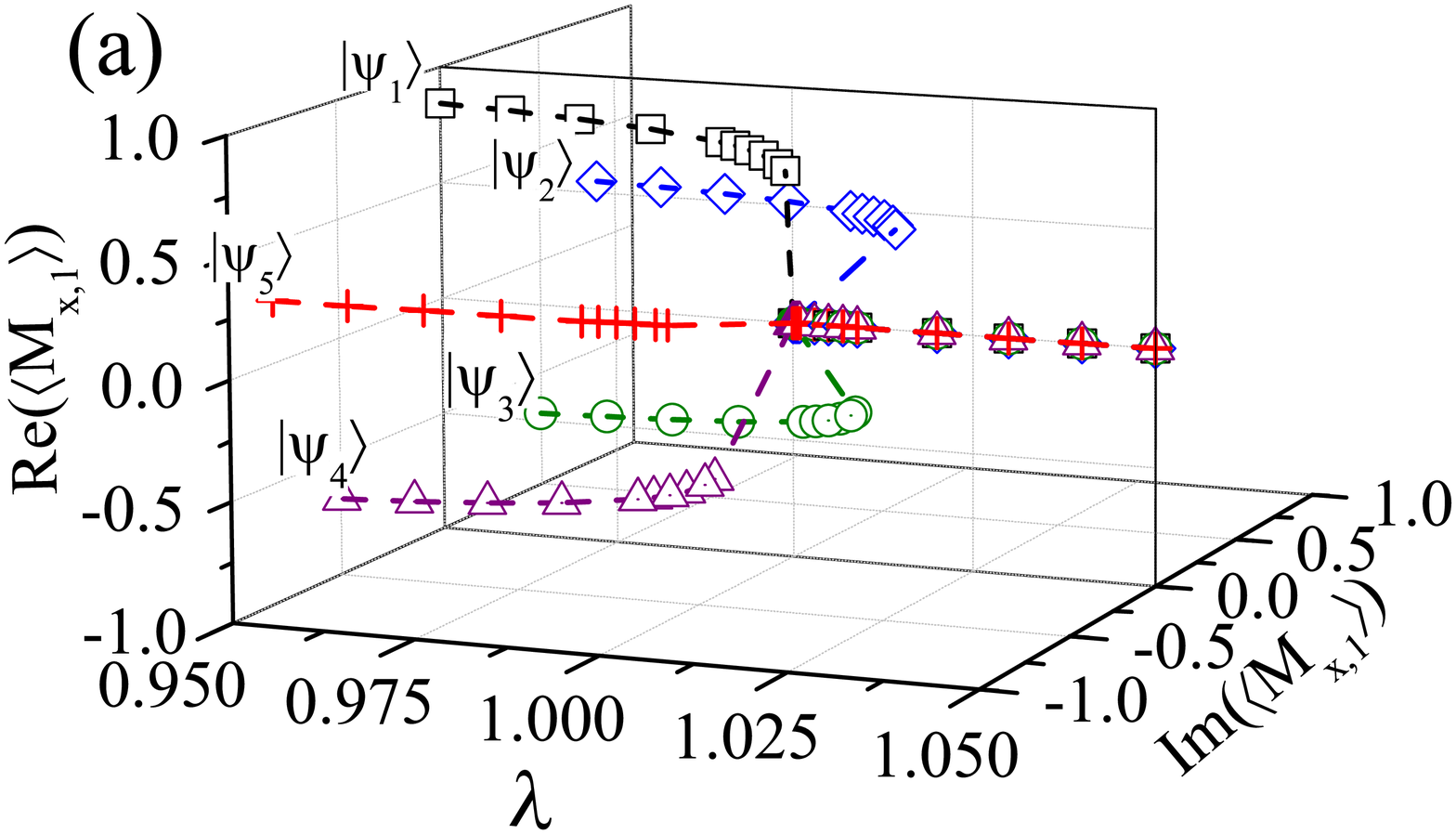}
  \vspace*{-0.5cm}
  \includegraphics[width=2.5in]{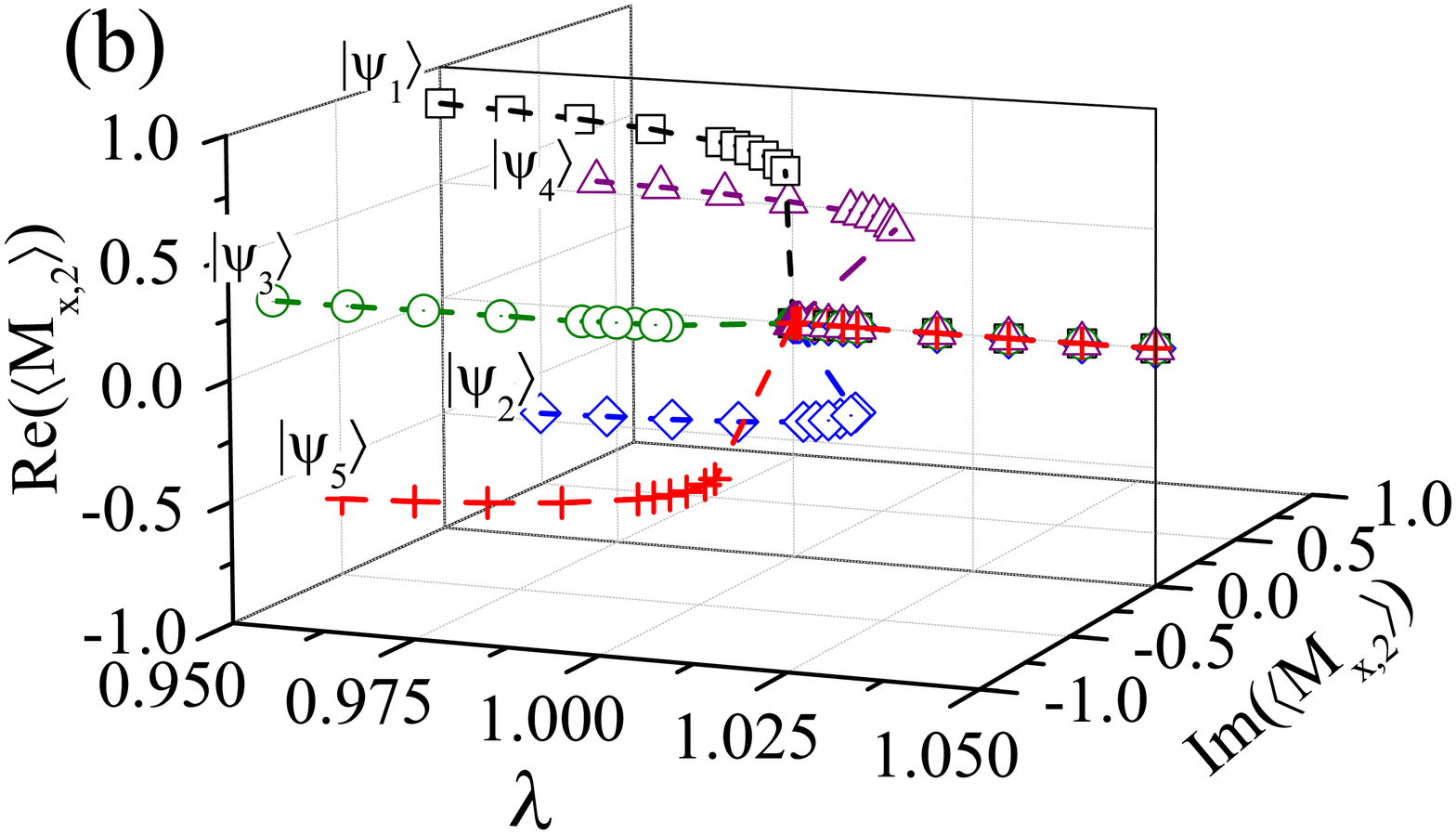}
  \vspace*{-0.5cm}
  \includegraphics[width=2.5in]{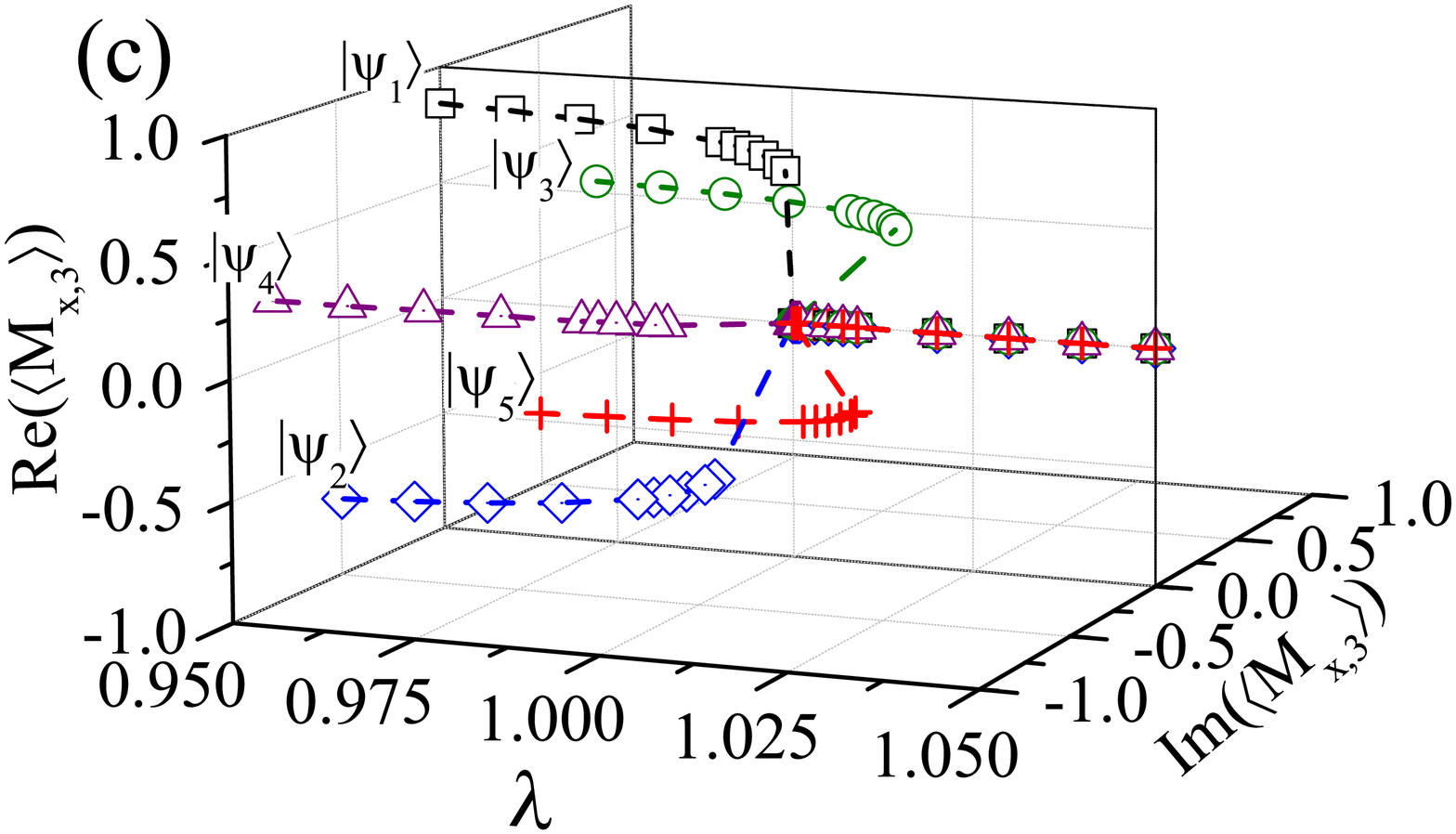}
  \includegraphics[width=2.5in]{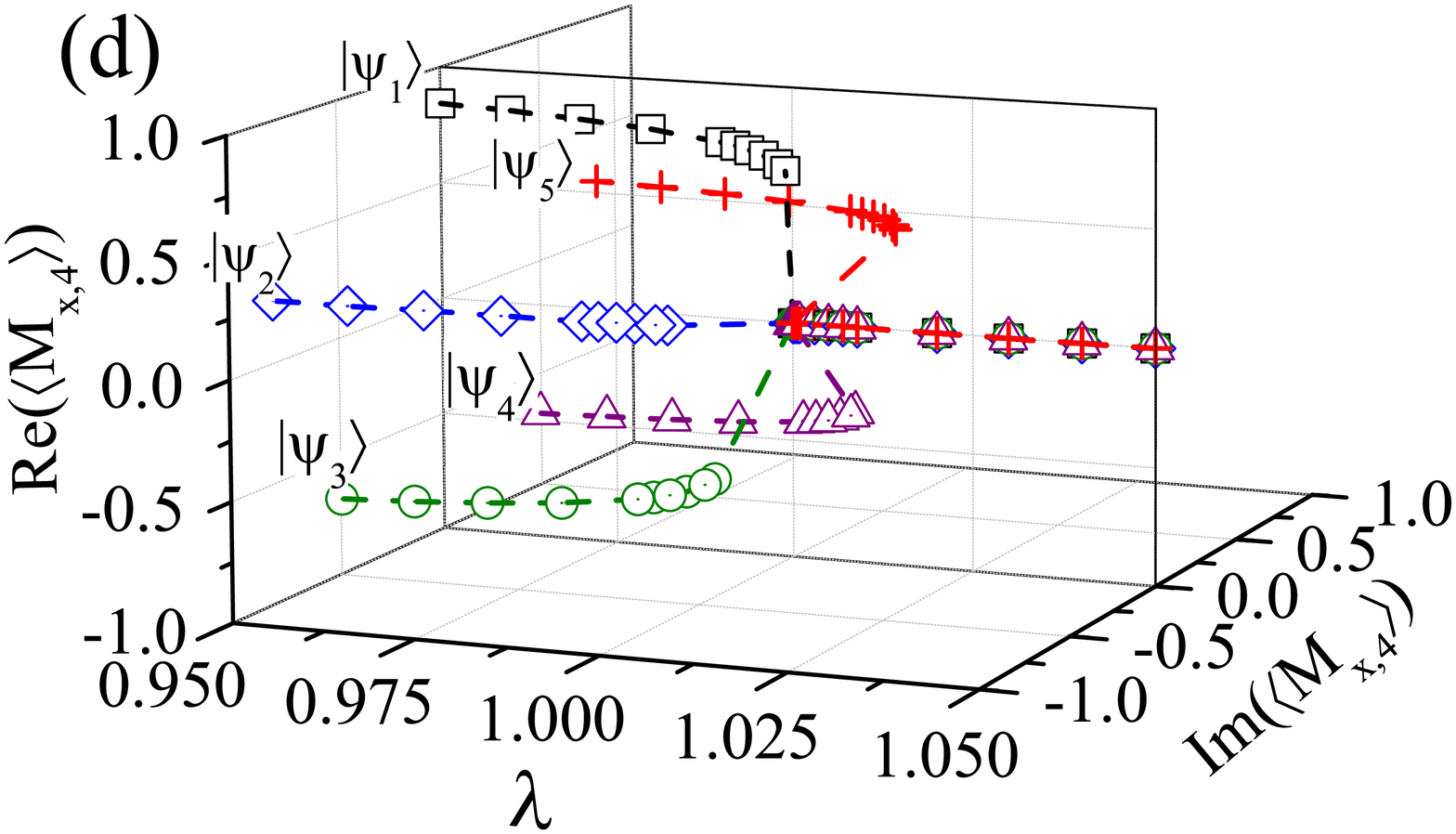}
 \end{center}
  \caption{(Color online) Magnetizations (a) $\langle M_{x,1}
 \rangle$, (b) $\langle M_{x,2} \rangle$,
 (c) $\langle M_{x,3} \rangle$,
 and (d) $\langle M_{x,4} \rangle$
 as a function of the transverse magnetic field $\lambda$
 for the $5$-state Potts model.
 For the broken symmetry phase,
 the magnetizations are given from each of the five degenerate groundstates.
 The numerical critical point is the exact value $\lambda_c=1$.
 }\label{fig6}
 \end{figure}
%%%%%%%%                one coloum   %%%%%%%%%%%%%%%%%%%%%%%

\end{widetext}

 Next, we consider
 the magnetizations for the $4$-state Potts
 model, i.e., $q=4$.
 In Fig. \ref{fig5}, we plot the magnetization (a) $\langle M_{x,1}
 \rangle$, (b) $\langle M_{x,2}\rangle$, and (c) $\langle M_{x,3}\rangle$
 as a function of the traverse magnetic field $\lambda$.
 The numerical critical point is $\lambda_c=1.0004$.
 All the magnetizations show that the phase transition
 is a continuous phase transition.
 Similar to the case $q=3$, all the absolute values of the magnetizations
 have the same values at a given magnetic field
 and
 the magnetizations in the complex
 magnetization plane have a relation between them under a rotation,
 which is characterized by the value $\omega_4 = \exp[2\pi i/4]$.
 In Fig. \ref{fig5}, we observe that,
 for a given magnetic field $\lambda < \lambda_c$,
 there are the relations between the magnetizations as
 $\langle M_{x,1}\rangle_1=\omega^{-1}_4 \langle M_{x,1}\rangle_2
 =\omega^{-2}_4 \langle M_{x,1}\rangle_3
 =\omega^{-3}_4 \langle M_{x,1}\rangle_4$ from (a),
 $\langle M_{x,2}\rangle_1=\omega^{-2}_4 \langle M_{x,2}\rangle_2
 =\omega^{-4}_4 \langle M_{x,2}\rangle_3
 =\omega^{-6}_4 \langle M_{x,2}\rangle_4$ from (b),
 and
 $\langle M_{x,3}\rangle_1=\omega^{-3}_4 \langle M_{x,3}\rangle_2
 =\omega^{-6}_4 \langle M_{x,3}\rangle_3
 =\omega^{-9}_4 \langle M_{x,3}\rangle_4$ from (c).
 Also, for each groundstate wavefunction, the magnetizations
 have the relations as
 $\langle M_{x,1}\rangle_1=\langle M_{x,2}\rangle_1=\langle M_{x,3}\rangle_1$,
 $\langle M_{x,1}\rangle_2=\omega^{-1}_4 \langle M_{x,2}\rangle_2
 =\omega^{-2}_4 \langle M_{x,3}\rangle_2$,
 $\langle M_{x,1}\rangle_3=\omega^{-2}_4 \langle M_{x,2}\rangle_3
 =\omega^{-4}_4 \langle M_{x,3}\rangle_3$,
and
 $\langle M_{x,1}\rangle_4=\omega^{-3}_4 \langle M_{x,2}\rangle_4
 =\omega^{-6}_4 \langle M_{x,3}\rangle_4$.
 These results show that, in the complex magnetization plane,
 the rotations between the magnetizations
 for a given magnetic field
 are determined by the characteristic rotation angles $\theta = 0$, $2\pi/4$,
 $4\pi/4$, and $6\pi/4$, i.e.,
 $\langle M_{x,p}\rangle_m=g_4 \langle M_{x,p'}\rangle_{m'}$ with
 $g_4 \in \{ I, \omega_4, \omega^2_4, \omega^3_4 \}$.
 Also, the degenerate groundstates give the same values for the
 $z$-component magnetizations, i.e., $\langle M_{z}\rangle_1=\langle M_{z}\rangle_2
  =\langle M_{z}\rangle_3=\langle M_{z}\rangle_4$.

 From the discussions for the $3$-state and $4$-state Potts models,
 one may refer a general relation between the magnetizations for any
 $q$-state Potts model. Actually, for any $q$, we find the relations
\begin{subequations}
   \label{eq6}
\begin{eqnarray}
 \omega^{p(1-m)}_q  \langle M_{x,p} \rangle_m
   &=& \omega^{p'(1-m')}_q\langle M_{x,p'} \rangle_{m'}, \\
    \langle M_{z} \rangle_m
   &=& \langle M_{z} \rangle_{m'},
\end{eqnarray}
\end{subequations}
 where $\omega_q = \exp[2\pi i/q]$.
 When one calculate the magnetizations of the operator $M_{x,p}$
 with different wavefunctions,
 the relations in Eq. (\ref{eq6}) reduce
 to $\langle M_{x,p} \rangle_m = \omega^{p(m-m')}_q \langle M_{x,p}
 \rangle_{m'}$ that satisfies the relations of the magnetizations
 in Figs. \ref{fig4} and \ref{fig5}.
 Also, if one choose a wavefunction,
 the magnetizations of the operators $M_{x,1}$, $\cdots$, $M_{x,q-2}$, and
 $M_{x,q-1}$ have the relations
 $\langle M_{x,p} \rangle_m = \omega^{(p'-p)(1-m)}_q \langle M_{x,p'}
 \rangle_m $ reduced from the relations in Eq. (\ref{eq6}).
 Furthermore,
 these results show that, in the complex magnetization plane,
 the rotations between the magnetizations
 for a given magnetic field
 are determined by the characteristic rotation angles $\theta = 0$, $2\pi/q$,
 $4\pi/q$, $6\pi/q$, $\cdots$, and $2(q-1)\pi/q$.
 Thus, Eq. (\ref{eq6}) can be rewritten as
\begin{subequations}
 \label{eq7}
\begin{eqnarray}
 \langle M_{x,p}\rangle_m &=& g_q \langle M_{x,p'}\rangle_{m'} ,
 \\
 g_q  %= \omega^{-p(1-a)}_q \omega^{p(1-a)}_q
  &\in& \{ I, \omega_q, \omega^2_q, \cdots, \omega^{q-1}_q \}.
\end{eqnarray}
\end{subequations}
 Then, Eq. (\ref{eq7}) shows clearly that
 the $q$-state Potts model has
 the discrete symmetry group $Z_q$ consisting of $q$ elements.
 For $q=2$, the $q$-state Potts model becomes
 the quantum Ising model that has a doubly degenerate groundstate
 due to a $Z_2$ symmetry.
 From Eq. (\ref{eq7}), one can easily  confirm
 $\langle M_{x}\rangle_1= - \langle M_{x}\rangle_2$
 because of $g_2 \in \{ I, \omega_2\}$ with $\omega_2=\exp[\pi i]$.
 Also,
 the $q$-state Potts Hamiltonian in Eq. (\ref{ham})
 is invariant with respect to the $q$ way transformations, i.e.,
 $U_m H_q U^\dagger_m = H_q$ for $m \in [1, q]$, which implies
 that the system has the $q$-fold degenerate groundstates for the broken
 symmetry phase according to the spontaneous symmetry breaking mechanism.
 The transformations are  given as
\begin{equation}
 U_m : \left\{
        \begin{array}{ccc}
        M_{x,p} & \rightarrow & \big(\omega^{p}_q\big)^{m-1} M_{x,p}
        \\
         M_z & \rightarrow & M_z
 \end{array} \right. ,
 \end{equation}
 where  $\omega_q =\exp[2\pi i/q]$.
 Although the $q$-state Potts Hamiltonian
 remains invariant under the full $q$ transformations, for the $Z_q$ broken symmetry
 phase,
 the $q$ order states described by
 the $q$ degenerate groundstates are invariant under only the subgroup of
 the $Z_q$ symmetry group.
 Obviously, Eqs. (\ref{eq6}) and (\ref{eq7}) show the relations between the order
 parameters of the $q$ equivalent ordered states under the $q$
 transformations.
 As a consequence, it is shown that,
{\it  from the degenerate groundstates,
 one can determine the order parameters as the magnetizations
 and their specification of how the order parameters transform
 under the symmetry group $Z_q$.}

 To make clearer
 the general relation of the magnetizations in Eqs. (\ref{eq6}) and (\ref{eq7}),
 let us consider the $5$-state Potts model.
 In Fig. \ref{fig6},
 we plot the magnetization (a) $\langle M_{x,1}
 \rangle$, (b) $\langle M_{x,2}\rangle$, (c) $\langle
 M_{x,3}\rangle$,
 and (d) $\langle M_{x,4}\rangle$
 as a function of the traverse magnetic field $\lambda$.
 The numerical critical point is obtained as the exact value $\lambda_c=1$.
 Also, all the magnetizations shows that the phase transition
 is a discontinues phase transition.
 Similar to the cases $q=3$ and $q=4$,
 all the magnetizations
 have the same values at a given magnetic field.
 and the magnetizations in the complex
 magnetization plane have a relation between them under a rotation,
 which is characterized by the value $\omega_5 = \exp[2\pi i/5]$.
 From Figs. \ref{fig6} (a)-(d), the relations between the magnetizations
 agree with those in Eqs. (\ref{eq6}) and (\ref{eq7}).

\section{Critical exponents, Central charge, and universality }

%%%%%%%%%%%%%%%%%%%%%%%%%%%%%%%%%%%%%%%%%%%%%%%%%%%%%%%%%%%%%%%%%%%%%%%%%%%
 \begin{center}
 \begin{table}[b]
 \caption{Critical exponents and central charge for $4$-state Potts model}
 \begin{tabular}{c||c|c|c|c|c|c|c}
 \hline \hline
 ${\rm  Exponents}$&$ \alpha $&$ \beta $&
 $ \gamma $&$ \delta $&$ \nu $&$ \eta $&$ c $\\
 \hline \hline
 $ {\rm Exact\footnote{The exact vaules of the critical exponents and central charge are taken
 from Refs.~\onlinecite{Wu} and ~\onlinecite{Baillie}. } }
  $&$ 2/3 $&$ 1/12 $ &$ 7/6 $ & $ 15 $ &$ 2/3 $&$ 1/4 $&$ 1 $\\
 $ {\rm iMPS} $&$ $&$ 0.0843 $&$ 1.0718 $& $  $&$ 0.6300 $&$ 0.2510 $&$ 0.9803 $\\
 \hline \hline
 \end{tabular}
 \label{table}
 \end{table}
 \end{center}
%%%%%%%%%%%%%%%%%%%%%%%%%%%%%%%%%%%%%%%%%%%%%%%%%%%%%%%%%%%%%%%%%%%%

%%

 As is known, for $q\leq4$, the quantum phase transitions are a continuous
 phase transition. If $q > 4$,  discontinuous (first order) phase transitions
 occur in the $q$-state Potts model.
 In this sense, then, the critical $q$ is $q_c=4$.
 In the section, we will study the critical exponents of $q_c=4$
 based on the iMPS groundstate wavefunctions. Actually, all the
 degenerate groundstates for the broken symmetry phase give a same
 exponents as it should be. In addition, in order to calculate
 a central charge, we discuss the von Neumann entropy for $q=3$, $4$,
 and $5$. In the Table \ref{table}, for the $4$-state Potts model,
 the critical exponents and the critical charge
 from the our iMPS results are compared with their exact values.

%%%%%%%%%%%%%%%%%%%%%%%%%%%%%%%%%%%%%%%%%%%%%%%%%%%%%%%%%%%%%%%%%%%%%
 \begin{figure}
 \begin{center}
 \includegraphics[width=3.in]{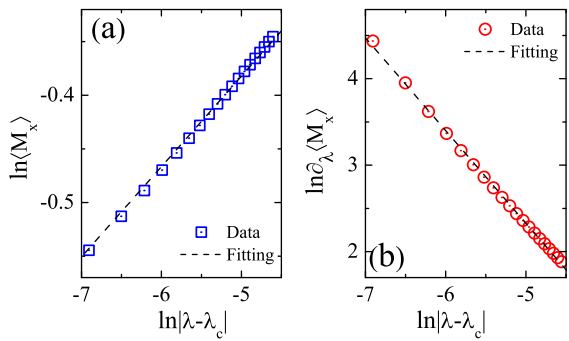}
 \end{center}
 \caption{(Color online) (a) Magnetization $\langle M_x \rangle$
 and (b) susceptibility $\partial_\lambda \langle M_x \rangle $ as a function
 of $|\lambda-\lambda_c|$ near the critical point $\lambda_c=1.0004$
 for $4$-state Potts model.
 }\label{fig7}
 \end{figure}
%%%%%%%%%%%%%%%%%%%%%%%%%%%%%%%%%%%%%%%%%%%%%%%%%%%%%%%%%%%%%%%%%%%%%%%%5

{\it Critical exponents}.$-$
 In our iMPS approach, we obtain the $4$-fold degenerate
 groundstates
 at zero temperature.
 The critical exponents $\alpha$ for specific heat and $\delta$
 for the field dependence of the magnetization at the critical temperature
 cannot be calculated.
 In the following, we discuss the four exponents, i.e.,
 $\beta$, $\gamma$, $\eta$, and $\nu$.
 Thus, we start with the magnetization near the critical
 point $\lambda_c=1.0004$.
 In Fig. \ref{fig7}, (a) the magnetization $\langle M_x\rangle$ and (b) the
 susceptibility $\partial_\lambda \langle M_x\rangle$
 are plotted as a function of $|\lambda-\lambda_c|$
 in the log-log plot.
 It is shown clearly that both the magnetization and the
 susceptibility can be described by power laws, i.e.,
 $\langle M_x\rangle \propto
 |\lambda-\lambda_c|^\beta$
 and $\partial_\lambda \langle M_x\rangle
 \propto |\lambda-\lambda_c|^{-\gamma}$
 with their characteristic exponents $\beta$ and $\gamma$, respectively.
 Thus, the fitting function is chosen to
  be $\ln \langle M_x \rangle = \beta\ln |\lambda-\lambda_c|+\beta_0$
  for the magnetization
  and
 $\ln \partial_\lambda \langle M_x \rangle =-\gamma
  \ln |\lambda-\lambda_c|+\gamma_0$
 for the susceptibility with the fitting constants
 $\beta_0$ and $\gamma_0$, respectively.
 From the numerical fittings,
 we obtain (a) $\beta=0.0843$ and
  $\beta_0=0.0394$ for the magnetization
  and (b) $\gamma=1.0718$ and $\gamma_0=-3.0308$
  for the susceptibility.
 The fitted critical exponents
 $\beta=0.0843$ and
 $\gamma=1.0718$ are
  quite close to the exact values\cite{Wu}
 $\beta=1/12\ (= 0.0833)$ and
 $\gamma=7/6\ (= 1.1667)$, respectively.

%%%%%%%%%%%%%%%%%%%%%%%%%%%%%%%%%%%%%%%%%%%%%%%%%%%%%%%%%%%%%%%%%%%%%%%%%
  \begin{figure}
 \begin{center}
  \includegraphics[width=3.in]{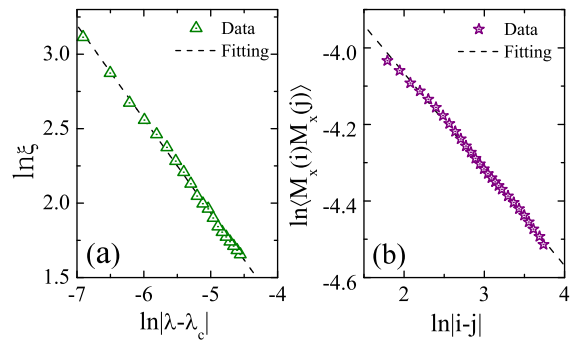}
 \end{center}
 \caption{(Color online)
 (a) Correlation length $\xi$ as a function
 of $|\lambda-\lambda_c|$ for the $4$-state Potts model.
 (b) Correlation $\langle M_{x}(i)M_{x}(j)\rangle$
 as a function of the lattice site distance
 $|i-j|$ at the critical point $\lambda_c=1.0004$.
 }\label{fig8}
 \end{figure}
%%%%%%%%%%%%%%%%%%%%%%%%%%%%%%%%%%%%%%%%%%%%%%%%%%%%%%%%%%%%%%%%%%%%%%%%%

 In Fig. \ref{fig8}, we plot (a) the correlation length $\xi$
 as a function of $|\lambda-\lambda_c|$ and (b) the correlation
 $\langle M_{x}(i)M_{x}(j)\rangle$ as a function of the lattice site
 distance $|i-j|$ at the critical point $\lambda_c=1.0004$.
 It is shown clearly that both the correlation length and the
 correlation at the critical point $\lambda_c$ can be described by power laws, i.e.,
 $\xi \propto
 |\lambda-\lambda_c|^{-\nu}$
 and $\langle M_{x}(i)M_{x}(j)\rangle
 \propto |\lambda-\lambda_c|^{-\eta}$
 with their characteristic exponents $\nu$ and $\eta$, respectively.
 We choose the fitting functions as
  $\ln \xi = -\nu\ln |\lambda-\lambda_c|+\nu_0$
  for the correlation length
  and
 $\ln \langle M_{x}(i)M_{x}(j)\rangle =-\eta
  \ln |i-j|+\eta_0$
 for the correlation at the critical point with the fitting constants
 $\nu_0$ and $\eta_0$, respectively.
 From the numerical fittings,
 we obtain (a) $\nu=0.6300$ and
  $\nu_0=-1.2153$ for the correlation length
  and (b) $\eta=0.2510$ and $\eta_0=-3.5644$
  for the correlation.
 The fitted critical exponents
 $\nu=0.6300$ and
 $\eta=0.2510$ are
  quite close to the exact values\cite{Wu}
 $\nu=2/3$ and
 $\eta=1/4$, respectively.

 {\it von Neumann entropy and central charge}.$-$
 In our iMPS approach, the von Neumann entropy $S$ can be
 directly evaluated by
 the elements of the diagonal matrix $\lambda^{[i]}_{\alpha_i}$
 that are the Schmidt decomposition coefficients of the bipartition
 between the semi-infinite chains $L(-\infty,...,i)$ and
 $R(i+1,...,\infty)$.
 This implies that Eq. (\ref{wave}) can be rewritten by
 $|\Psi\rangle = \sum^\chi_{\alpha=1} \lambda_{\alpha}
 |\psi^L_\alpha\rangle |\psi^R_\alpha \rangle$,
 where $|\psi^L_\alpha \rangle$ and $|\psi^R_\alpha \rangle$ are the Schmidt bases
 for the semi-infinite chains $L(-\infty,...,i)$ and
 $R(i+1,...,\infty)$, respectively.
 For the bipartition, then,
 the von Neumann entropy $S$ can be defined as \cite{Bennett}
 $S= - {\rm Tr} [\varrho_L \log
 \varrho_L] = - {\rm Tr} [\varrho_R \log \varrho_R]$,
 where $\varrho_L =
 {\rm Tr}_R\,  \varrho$ and $\varrho_R={\rm Tr}_L\,  \varrho$ are the reduced density
 matrices of the subsystems $L$ and $R$, respectively, with the density
 matrix $\varrho=|\Psi \rangle \langle \Psi|$.
 For the semi-infinite chains $L$ and $R$ in the iMPS representation,
 the von Neumann entropy $S$ calculated by
  $S = -\sum_{\alpha=1}^{\chi} \lambda^2_\alpha \log
  \lambda^2_\alpha$.
 Actually,
 the von Neumann entropy as one of the quantum entanglement measures
 have been proposed as a general indicator to determine and
 characterize quantum phase transitions\cite{Korepin,entropy}.
 Also,
 the logarithmic scaling of the von Neumann entropy
 was conformed to exhibit conformal invariance
 and the scaling is governed by a
 universal factor\cite{Tagliacozzo,Pollmann,Zanardi}, i.e., a central charge $c$ of the
 associated conformal field theory.
 In the iMPS approach, for a continuous phase transition,
 the diverging entanglement at a quantum
 critical point
 gives simple scaling relations\cite{Tagliacozzo} for (i) the von Neumann entropy
  $S$ and (ii) a correlation length $\xi_v$ with respect to
 the truncation dimension $\chi$ as
 $ S \sim \frac {c \kappa} {6} \log \chi $
 and  $\xi_v \sim  A \chi^{\,\kappa}$,
 where $c$ is a central charge, $\kappa$ is
 a so-called finite-entanglement scaling exponent, and $A$ is a constant.
 By using the relations, a central charge can be
 obtained numerically at a critical point.

%%%%%%%%%%%%%%%%%%%%%%%%%%%%%%%%%%%%%%%%%%%%%%%%%%%%%%%%%%%%%%%%%%%%%%
 \begin{figure}
 \begin{center}
 \includegraphics[width=3.in]{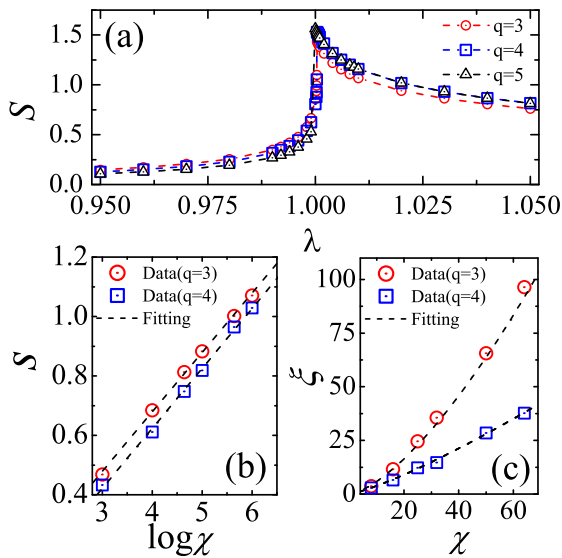}
 \end{center}
 \caption{(Color online) (a) Groundstate von Neumann entropies
 for the $3$-, $4$-, and $5$-state Potts models
 as a function of the transverse magnetic field $\lambda$.
  (b) Von Neumann entropies $S$ and (c) Correlation lengthes $\xi_v$
  for $q=3$ and $q=4$
  with respect to the truncation dimension $\chi$
  at the exact critical point $\lambda_c=1$.
 }\label{fig9}
 \end{figure}
%%%%%%%%%%%%%%%%%%%%%%%%%%%%%%%%%%%%%%%%%%%%%%%%%%%%%%%%%%%%%%%%%%%%%%%%%%%%%%%%
%
 In Fig. \ref{fig9} (a), we plot the von Neumann entropies
 as a function of the transverse magnetic field $\lambda$
 for the $3$-, $4$-, and $5$-state Potts models with the truncation dimension $\chi=32$.
 In the entropies, there are singular points at $\lambda_c =1.0004$ for $q=3$ and $4$,
 and $\lambda_c =1$ for $q=5$, which are consistent with the critical
 points from the multiple bifurcation points in Fig.
 \ref{fig3} and from the magnetizations as the order
 parameters in Figs. \ref{fig4}, \ref{fig5}, and \ref{fig6}.
 It is shown clearly that the von Neumann entropies
 capture the phase transitions.
 The (dis-)continuity of the von Neumann entropies for $q=3$ and $q=4$ ($q=5$) indicates
 a (dis-)continuous phase transition between the broken symmetry phases and the symmetry phases.
 However,
 the von Neumann entropies for the different degenerate
 groundstates give the same values,
 which implies that the von Neumann entropy
 cannot distinguish the different degenerate
 groundstates in the broken symmetry phases.

 For $q=3$ and $q=4$, in Figs. \ref{fig9} (b) and (c), we plot
 (b) the von Neumann entropy and (c) the correlation length $\xi_v$
 as a function of the truncation
 dimension $\chi$ at the critical points $\lambda_{c} =1$.
 Here, the truncation dimensions are taken as $\chi=8$, $16$, $25$,
 $32$, $50$, and $64$.
 It is shown that both the von
 Neumann entropy $S$ and the correlation length $\xi_v$
 diverge as the truncation dimension $\chi$ increases.
 In order to obtain the central charges,
 we use the numerical fitting functions, i.e.,
 $S_q(\chi) = a_q + b_q \log \chi$ and $\xi_{v,q}(\chi) = A_q \chi^{\,\kappa_q}$.
 Numerically,
 the constants of the von Neumann entropies are fitted as
  $a_3= -0.1215$ and $b_3=0.2000$ for $q=3$
  and
  $a_4= -0.1830$ and $b_4= 0.2016$ for $q=4$.
 The power-law fittings on the correlation lengthes $\xi_{v,q}$
 give the fitting constants as
 $A_3=0.2060$ and
 $\kappa_3=1.4775$ for $q=3$.
 and
 $A_4=0.2243$ and
 $\kappa_4=1.2340$ for $q=4$.
  As a result,
 from the $\kappa_q$ and $b_q$,
 the central charges are obtained as
 $c_3=0.8121$ for $q=3$
 and
 $c_4=0.9803$ for $q=4$,
 which are quite close to the exact values\cite{Baillie} $c=4/5$ for $q=3$
 and $c=1$ for $q=4$, respectively.

\section{Summary}
 We have investigated how to obtain degenerate groundstates
 by using a quantum fidelity.
 To do this, we have introduced the quantum fidelity
 between the degenerate groundstates and an arbitrary reference
 state.
 The distinguished degenerate groundstate wavefunctions
 allows naturally to detect a quantum phase transition that
 is indicated by a multiple bifurcation point of the quantum
 fidelity per lattice site.
 Furthermore, as the complete description of an order phase,
 it is possible to classify how the order parameters calculated
 from the degenerate groundstates transform under the subgroup
 of a symmetry group of Hamiltonian.
 As an example, the $q$-state Potts model has been investigated
 by employing the iMPS with the iTEBD.
 We have obtained the $q$-fold degenerated groundstates explicitly.
 A general relation between the magnetizations calculated from the degenerate groundstates
 is obtained to show
 the spontaneous symmetry breaking of the $Z_q$ symmetry group
 in the $q$-state Potts model.
 In addition,
 the critical exponents and the central charges
 are directly calculated from the degenerate groundstates,
 which is shown that the iMPS results
 are quite close to the exact values.

\section*{Acknowledgements}
 YHS thanks Bo Li for helpful discussions about the critical exponents.
 We thank Huan-Qiang Zhou for helpful discussions to inspire and encourage us
 to complete this work.
 The work was supported by the National Natural Science Foundation of
 China (Grant No. 11174375).


\begin{thebibliography}{1}
%
 \bibitem{QPT}
 S. Sachdev, {\it Quantum Phase Transitions} (Cambridge University, Cambridge, 1999).
%
 \bibitem{Chaikin}
 P. M. Chaikin and T. C. Lubensky,
 {\it Pinciples of Condensed Matter Physics} (Cambridge University, Cambridge, 1995).
%
 \bibitem{Landau}
 L. D. Landau and E. M. Lifshitz, {\it Statistical Physics} (Pergamon, New York, 1958).
%
 \bibitem{Zhou1}
 H.-Q. Zhou and J. P. Barjaktarevi\v{c},  J. Phys. A: Math. Theor. \textbf{41}, 412001 (2008);
 H.-Q. Zhou, J.-H. Zhao, and B. Li, J. Phys. A: Math. Theor. \textbf{41}, 492002
 (2008).
%
 \bibitem{Zanardi}
 P. Zanardi and N. Paunkovi\'{c}, Phys. Rev. E \textbf{74}, 031123 (2006).

 \bibitem{Rams}
 M. M. Rams and B. Damski, Phys. Rev. Lett. \textbf{106}, 055701 (2011).
%
 \bibitem{Mukherjee}
 V. Mukherjee and A. Dutta, Phys. Rev. B \textbf{83}, 214302 (2011).
%
 \bibitem{Liu}
  H.-Q. Zhou, arXiv:0704.2945;
  E. Eriksson and H. Johannesson, Phys. Rev. A \textbf{79}, 060301(R) (2009);
 Z. Wang, T. Ma, S.-J. Gu, and H.-Q. Lin, Phys. Rev. A \textbf{81}, 062350 (2010);
 J.-H. Liu, Q.-Q. Shi, J.-H. Zhao, and H.-Q. Zhou, J. Phys. A: Math. Theor. \textbf{44}, 495302 (2011).
%

%
 \bibitem{Gu}
 S.-J. Gu, Int. J. Mod. Phys. B \textbf{24}, 4371 (2010).
%
 \bibitem{Xiao}
 X. Wang, Z. Sun, and Z. D. Wang, Phys. Rev. A \textbf{79}, 012105 (2009).

%
 \bibitem{Wang}
 H.-L. Wang, J.-H. Zhao, B. Li, and H.-Q. Zhou, J. Stat. Mech., L10001 (2011).
%
 \bibitem{mps}
 M. Fannes, B. Nachtergaele, and R. F. Werner, Comm. Math. Phys, \textbf{144}, 3 (1994);
 S. \"{O}stlund and S. Rommer. Phys. Rev. Lett. \textbf{75}, 3537 (1995);
 D. Perez-Garcia, F. Verstraete, M. M. Wolf, and J. I. Cirac, Quantum Inf. Comput. \textbf{7}, 401 (2007).
%
 \bibitem{DMRG}
 S. R. White, Phys. Rev. Lett. \textbf{69}, 2863 (1992);
 S. R. White, Phys. Rev. B. \textbf{48}, 10345 (1993);
 U. Schollw\"ock, Rev. Mod. Phys. \textbf{77}, 259 (2005).
%
 \bibitem{Vidal}
 G. Vidal, Phys. Rev. Lett. \textbf{91}, 147902 (2003);
 G. Vidal, Phys. Rev. Lett. \textbf{98}, 070201 (2007).
%

%
 \bibitem{Zhou2}
 H.-Q. Zhou, Roman Or\'{u}s, and G. Vidal, Phys. Rev. Lett. \textbf{100}, 080601 (2008).
%
 \bibitem{Zhao}
 J.-H. Zhao, H.-L. Wang, B. Li, and H.-Q. Zhou, Phys. Rev. E \textbf{82}, 061127
 (2010).
 \bibitem{Bi}
 S.-H. Li, H.-L. Wang, Q.-Q. Shi, and H.-Q. Zhou, arXiv:1105.3008;
 H.-L. Wang, Y.-W. Dai, B.-Q. Hu, and H.-Q. Zhou, Phys. Lett. A \textbf{375}, 4045
 (2011).

 \bibitem{Dai}
 Y.-W. Dai, B.-Q. Hu, J.-H. Zhao, and H.-Q. Zhou, J. Phys. A: Math. Theor. \textbf{43}, 372001 (2010).
%
 \bibitem{Potts}
 R. B. Potts, Proc. Cambridge Phil. Soc. \textbf{48}, 106 (1952).
%
 \bibitem{Wu}
 F. Y. Wu, Rev. Mod. Phys. \textbf{54}, 235 (1982);
 R. J. Baxter, {\it Exactly Solved Models in Statistical Mechanics} (Academic, London, 1982);
 P. P. Martin, {\it Potts Models and Related Problems in Statistical Mechanics} (World Scientific, Singapore, 1991).
%
 \bibitem{Murg}
 V. Murg, F. Verstraete, and J. I. Cirac,  Phys. Rev. A \textbf{75}, 033605 (2007).
%
 \bibitem{suzuki}
 M. Suzuki, Phys. Lett. A \textbf{146}, 319 (1990).
%
 \bibitem{Baxter2}
 R. J. Baxter, J. Phys. C \textbf{6}, L445 (1973);
 R. J. Baxter, J. Phys. A \textbf{15}, 3329 (1982).
%
 \bibitem{Nijs}
 M. P. M. Nijs, J. Phys. A \textbf{12}, 1857 (1979);
 M. P. M. Nijs, Phys. Rev. B \textbf{27}, 1674 (1983).
%
 \bibitem{Pearson}
 R. B. Pearson, Phys. Rev. B \textbf{22}, 2579 (1980).
%
 \bibitem{Black}
 J. Black and V. J. Emery, Phys. Rev. B \textbf{23}, 429 (1981).
%
 \bibitem{Caselle}
 M. Caselle, F. Gliozzi and S. Necco, J. Phys. A: Math. Gen. \textbf{34}, 351 (2001).
%
 \bibitem{Hove}
 J. Hove, J. Phys. A \textbf{38}, 10893 (2005).
%
 \bibitem{Michel}
 L. Michel, Rev. Mod. Phys. \textbf{52}, 617 (1980).
%
\bibitem{Baillie}
 C. F. Baillie and D. A. Johnston, Mod. Phys. Lett. A \textbf{7}, 1519 (1992).
%
 \bibitem{Bennett}
 C. H. Bennett, H. J. Bernstein, S. Popescu, and B. Schumacher, Phys. Rev. A \textbf{53}, 2046 (1996).
%
 \bibitem{Korepin}
 V. E. Korepin, Phys. Rev. Lett. \textbf{92}, 096402 (2004);
 P. Calabrese and J. Cardy, J. Stat. Mech.: Theory Exp., P06002 (2004).
%
 \bibitem{entropy}
 L. Amico, R. Fazio, A. Osterloh, and V. Vedral, Rev. Mod. Phys. \textbf{80}, 517 (2008).
%
 \bibitem{Tagliacozzo}
 L. Tagliacozzo, T. R. de Oliveira, S. Iblisdir, and J. I. Latorre, Phys. Rev. B \textbf{78}, 024410 (2008).
%
 \bibitem{Pollmann}
 F. Pollmann, S. Mukerjee, A. M. Turner, and J. E. Moore, Phys. Rev. Lett. \textbf{102}, 255701 (2009).



\end{thebibliography}
\end{document}